\documentclass[preprint, amsmath, amssymb, showkeys]{revtex4-1}

\usepackage{graphicx}
\usepackage{multirow}
\usepackage[bookmarks=false]{hyperref}

\begin{document}

\title{Defining and Identifying Sleeping Beauties in Science}

\author{Qing Ke}
\affiliation{Center for Complex Networks and Systems Research, School of Informatics and Computing, Indiana University, Bloomington, Indiana 47408, USA}
\author{Emilio Ferrara}
\affiliation{Center for Complex Networks and Systems Research, School of Informatics and Computing, Indiana University, Bloomington, Indiana 47408, USA}
\author{Filippo Radicchi}
\affiliation{Center for Complex Networks and Systems Research, School of Informatics and Computing, Indiana University, Bloomington, Indiana 47408, USA}
\author{Alessandro Flammini}
\affiliation{Center for Complex Networks and Systems Research, School of Informatics and Computing, Indiana University, Bloomington, Indiana 47408, USA}

\begin{abstract}
A Sleeping Beauty (SB) in science refers to a paper whose importance is
not recognized for several years after publication. Its citation history
exhibits a long hibernation period followed by a sudden spike of popularity.
Previous studies suggest a relative scarcity of SBs. The reliability of this
conclusion is, however, heavily dependent on identification methods based on
arbitrary threshold parameters for sleeping time and number of citations,
applied to small or monodisciplinary bibliographic datasets. Here we present
a systematic, large-scale, and multidisciplinary analysis of the SB
phenomenon in science. We introduce a parameter-free measure that quantifies
the extent to which a specific paper can be considered an SB. We apply our
method to 22 million scientific papers published in all disciplines of natural
and social sciences over a time span longer than a century. Our results reveal
that the SB phenomenon is not exceptional. There is a continuous spectrum of
delayed recognition where both the hibernation period and the awakening
intensity are taken into account. Although many cases of SBs can be identified by
looking at monodisciplinary bibliographic data, the SB phenomenon becomes much
more apparent with the analysis of multidisciplinary datasets, where we can
observe many examples of papers achieving delayed yet exceptional importance in
disciplines different from those where they were originally published. Our
analysis emphasizes a complex feature of citation dynamics that so far has
received little attention, and also provides empirical evidence against the use
of short-term citation metrics in the quantification of scientific impact.
\end{abstract}

\keywords{delayed recognition; Sleeping Beauty; bibliometrics}

\maketitle

\textbf{Significance}---Scientific papers have typically a finite lifetime: their
rate to attract citations achieves its maximum a few years after publication, and
then steadily declines. Previous studies pointed out the existence of a few
blatant exceptions: papers whose relevance has not been recognized for decades,
but then suddenly become highly influential and cited. The Einstein, Podolsky,
and Rosen ``paradox'' paper is an exemplar Sleeping Beauty. We study how common
Sleeping Beauties are in science. We introduce a quantity that captures both
the recognition intensity and the duration of the ``sleeping'' period, and show
that Sleeping Beauties are far from exceptional. The distribution of such
quantity is continuous and has power-law behavior, suggesting a common
mechanism behind delayed but intense recognition at all scales.

\vspace{25mm}

There is an increasing interest in understanding the dynamics underlying
scientific production and the evolution of
science~\cite{egghe1990introduction}. Seminal studies focused on scientific
collaboration networks~\cite{Newman-coauthor-04}, evolution of
disciplines~\cite{Sun-dynamics-13}, team science~\cite{Guimera-team-05,
Wuchty-team-07, Jones-team-08, Milojevic-team-14}, and citation-based scientific
impact~\cite{Radicchi-impact-08, Wang-impact-13, Uzzi-comb-13}. An important
issue at the core of many research efforts in science of science is
characterizing how papers attract citations during their
lifetime. Citations
can be regarded as the credit units that the scientific community attributes
to its research products. As such, they are at the basis of several quantitative
measures aimed at evaluating career trajectories of
scholars~\cite{hirsch2005index} and research performance of
institutions~\cite{kinney2007national, davis1984faculty}. They are also
increasingly used as evaluation criteria in very important contexts, such as
hiring, promotion, and tenure, funding decisions, or department and university
rankings~\cite{bornmann2006selecting, liu2005academic}. Several factors can
potentially affect the amount of citations accumulated by a
paper over time, including
its quality, timeliness, and potential to trigger further inquiries~\cite{Wang-impact-13},
the reputation of its authors~\cite{sarigol2014predicting, Petersen-reputation-14},
as well as its topic and
age~\cite{Radicchi-impact-08}.

Studies about fundamental mechanisms that drive citation dynamics started
already in the 1960s, when de~Solla~Price introduced the
cumulative advantage (CA) model to explain the emergence of power-law
citation distributions~\cite{Price-cumuadv-76}. CA essentially provisions that
the probability of a publication to attract a new citation is proportional to
the number of citations it already has. The criterion, now widely referred to
as preferential attachment, was recently popularized by Barab\'asi and
Albert~\cite{Barabasi-pa-99}, who proposed it as a general mechanism that
yields heterogeneous connectivity patterns in networks describing systems in
various domains~\cite{albert2002statistical, boccaletti2006complex}. Other
processes that effectively incorporate the CA mechanism have been proposed to
explain power-law citation distributions. Krapivsky and Redner, for example,
considered a redirection mechanism, where new papers copy with a certain
probability the citations of other papers~\cite{Krapivsky-redirect-01}.

An important effect not included in the CA mechanism is the fact that the
probability of receiving citations is time dependent. In the CA model, papers
continue to acquire citations independently of their age so that, on average,
older papers accumulate higher number of citations~\cite{Barabasi-pa-99,
Krapivsky-redirect-01, newman2009first}. However, it has been empirically
observed that the rate at which a paper accumulates citations decreases after
an initial growth period~\cite{hajra2004phase, hajra2005aging,
hajra2006modelling, wang2008measuring}. Recent studies about growing network
models include the aging of nodes as a key feature~\cite{hajra2004phase,
wang2008measuring, dorogovtsev2000evolution, dorogovtsev2001scaling,
zhu2003effect}. More recently, Wang et~al. developed a model that
includes, in addition to the CA and aging, an intuitive yet fundamental
ingredient: a fitness or quality parameter that accounts for the perceived
novelty and importance of individual papers~\cite{Wang-impact-13}.

In this work, we focus on the citation history of papers receiving an intense
but late recognition. Note that delayed recognition cannot be predicted by
current models for citation dynamics. All models, regardless of the number of
ingredients used, naturally lead to the so-called first-mover advantage,
according to which either papers start to accumulate citations in the early
stages of their lifetime or they will never be able to accumulate a significant
number of citations~\cite{newman2009first}. Back in the 1980s, Garfield provided
examples of articles with delayed recognition and suggested to use citation
data to identify them~\cite{Garfield-beauty-80, Garfield-beauty-89,
Garfield-beautyb-89, Garfield-beauty-90}. Through a broad literature search,
Gl\"anzel et al. gave an estimate for the occurrence of delayed recognition,
and highlighted a few shared features among lately recognized
papers~\cite{Glanzel-beauty-03}. The coinage of the term ``Sleeping Beauty"
(SB) in reference to papers with delayed recognition is due to
van~Raan~\cite{Raan-Beauty-04}. He proposed three dimensions along which
delayed recognition can be measured: (\emph{i}) length of sleep, i.e., the
duration of the ``sleeping period;'' (\emph{ii}) depth of sleep, i.e., the
average number of citations during the sleeping period; and (\emph{iii}) awake
intensity, i.e., the number of citations accumulated during 4 years after the
sleeping period. By combining these measures, he identified a few SB examples
occurred between $1980$ and $2000$.
These seminal studies suffer from two main limitations: (\emph{i}) the analyzed
datasets are very small, especially if compared to the size of the
bibliographic databases currently available; and (\emph{ii}) the definition and the
consequent identification of SBs are to the same extent arbitrary, and strongly
depend on the rules adopted. 
More recently, Redner analyzed a very
large dataset covering $110$ years of publications in
physics~\cite{Redner-110citation-2005}. Redner proposed a definition of
revived classic (or SB) for articles satisfying the three following
criteria: (\emph{i}) publication date antecedent 1961; (\emph{ii}) number of citations
larger than 250; and (\emph{iii}) ratio of the average citation age to publication age
greater than 0.7. Whereas Redner was able to overcome the first
limitation mentioned above, his study is still affected by an arbitrary
selection choice of top SBs, justified by the principle that SBs
represent exceptional events in science. In addition, Redner's analysis
has the limitation to be field specific, covering only publications
and citations within the realm of physics.

Here we perform an analysis on the SB phenomenon in science.
We propose a parameter-free approach to quantify how much a given paper
can be considered as an SB. We call this index ``beauty coefficient,"
denoted as $B$. By measuring $B$ for tens
of millions of publications in
multiple scientific disciplines
over an observation window longer than a century, we show that $B$ is
characterized by a heterogeneous but continuous distribution, with no natural
separation between papers with low, high, or even extreme values of $B$. Also,
we demonstrate that the empirical distributions of $B$ cannot be easily
reconciled with obvious baseline models for citation accumulation that are based
solely on CA or the reshuffling of citations. We introduce a simple
method to identify the awakening time of
SBs, i.e., the year when their 
citations burst.
The results indicate that many SBs become highly influential more than 50 years after their publication, far longer than typical time windows for measuring citation impact, corroborating recent studies on understanding the use of short time windows to approximate long-term citations~\cite{Bornmann-percentile-2013, Bornmann-improve-2014, Wang-window-2013}. 
We further show that the majority of papers exhibit a sudden decay of popularity after reaching the maximum number of yearly citations, independently of their $B$ values.
Our study points out that the SB phenomenon has two important multidisciplinary components.
First, particular disciplines, such as physics, chemistry, and mathematics, are able to produce top SBs at higher rates than other scientific fields.
Second, top SBs achieve delayed exceptional importance in disciplines different from those where they were originally published.
Based on these results, we believe that our study may pave the way to the identification of the complex dynamics that trigger the awakening mechanisms, shedding light on highly cited papers that follow nontraditional popularity trajectories.

\section{materials}
\subsection{Beauty coefficient} The beauty coefficient value $B$ for a given
paper is based on the comparison between its citation history and a reference
line that is determined only by its publication year, the maximum number of
citations received in a year (within a multi-year observation period), and the
year when such maximum is achieved. Given a paper, let us define $c_t$ as the
number of citations received in the $t$-th year after its
publication; $t$ indicates the age of the paper. Let us also assume that our
index $B$ is measured at time $t = T$, and that the paper receives its maximum
number $c_{t_m}$ of yearly citations at time $t_m \in [0,T]$.

\begin{figure}
\begin{center}
\includegraphics[trim=0mm 0mm 0mm 0mm, width=\columnwidth]{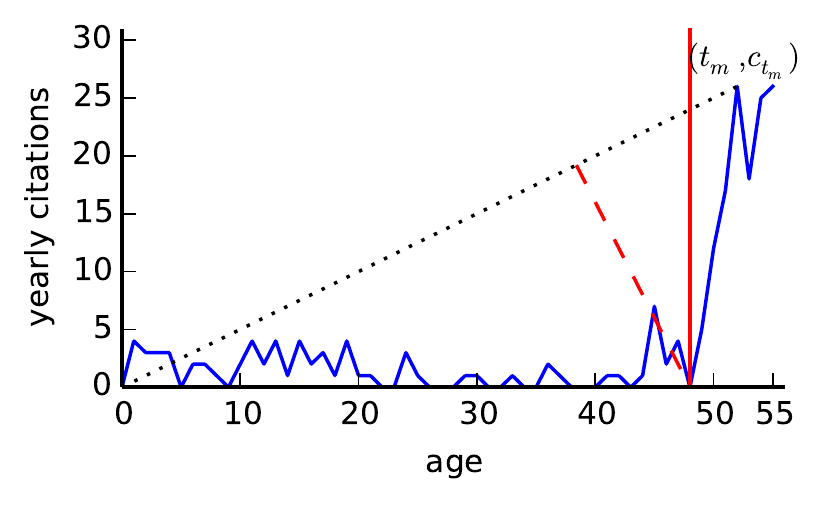}
\caption{\label{fig:beauty-cal-illu}Illustration of the definition of the beauty
coefficient $B$ (Eq.~\ref{eq:beauty}) and the awakening time $t_a$
(Eq.~\ref{eq:awake}) of a paper. The blue curve represents the number of
citations $c_t$ received by the paper at age $t$ (i.e., $t$ represents the
number of years since its publication). The black dotted line connecting the
points $(0,c_0)$ and $(t_m,c_{t_m})$ is the reference line $\ell_t$
(Eq.~\ref{eq:ref-line}) against which the citation history of the paper is
compared. The awakening time $t_a \leq t_m$ is defined as the age that
maximizes the distance from $(t,c_t)$ to the line $\ell_t$
(Eq.~\ref{eq:awake}), indicated by the red dashed line. The red vertical line
marks the awakening time $t_a$ calculated according to Eq.~\ref{eq:awake}.
The figure refers to the paper Phys Rev \textbf{95}(5):1154 (1954)~\cite{PhysRev.95.1154}.
}
\end{center}
\end{figure}

Consider the straight line $\ell_t$ that connects the points $(0,c_0)$ and
$(t_m,c_{t_m})$ in the time-citation plane (Fig.~\ref{fig:beauty-cal-illu}).
This line is described by the equation
\begin{equation}
\label{eq:ref-line}
\ell_t = \frac{c_{t_m}-c_0}{t_m} \cdot t + c_0, 
\end{equation}
where $\left( c_{t_m}-c_0 \right)/t_m$ is the slope of the line, and $c_0$ the
number of citations received by the paper in the year of its publication. For
each $t \leq t_m$, we then compute the ratio between $\ell_t - c_t$ and
$\max\{1,c_t\}$. Summing up the ratios from $t=0$ to $t=t_m$, the
beauty coefficient $B$ is defined as
\begin{equation} \label{eq:beauty}
B = \sum_{t=0}^{t_m} \frac{\frac{c_{t_m}-c_0}{t_m} \cdot t + c_0 - c_t}{\max\{1,c_t\}}.
\end{equation}
By definition, $B = 0$ for papers with $t_m = 0$. Papers with citations
growing linearly with time ($c_t = \ell_t$) have $B=0$. $B$ is non-positive
for papers whose citation trajectory $c_t$ is a concave function of time. Our
index $B$ has a number of desirable properties: (\emph{i}) $B$ can be computed for
any paper and does not rely on arbitrary thresholds on the sleeping period or
the awakening intensity, paving the way to treat the SB phenomenon not as just an
exception; (\emph{ii}) $B$ increases with both the length of the sleeping period and
the awakening intensity; (\emph{iii}) $B$ takes into account the entire citation
history in the time window $0 \leq t \leq t_m$; and (\emph{iv}) The denominator of
Eq.~\ref{eq:beauty} penalizes early citations so that, at
parity of total citations received, the later those citations are accumulated
the higher is the value of $B$.

\subsection{Awakening time} We now give a plausible definition of awakening time---the year when the abrupt
change in the accumulation of citations of SBs occurs. Being able to pinpoint
the awakening time may help identifying possible general trigger mechanisms
behind said change. For example, in SI Appendix we show that around the awakening
time, the SBs co-citation dynamics exhibit clear topical
patterns (SI Appendix, Fig.~S11)~\cite{Redner-110citation-2005}.
We define the awakening time $t_a$ as the time $t$ at which the distance $d_t$
between the point $(t, c_t)$ and the reference line $\ell_t$ reaches its
maximum:
\begin{equation}
\label{eq:awake}
t_a = \arg  \; \left\{ \max_{t \leq t_{m}} \; d_t \right\}.
\end{equation}
where $d_t$ is given by
\[
d_t = \frac{\left\vert (c_{t_m}-c_0) t - t_{m}c_t + t_{m}c_0 \right\vert}{\sqrt{(c_{t_m}-c_0)^2 + t_{m}^2}}.
\]

As we shall show, the above definition works well for limit cases where there
are no citations until the spike, and seems to well capture the qualitative
notion of awakening time when a strong SB-like behavior is present.

\subsection{Datasets} We use two datasets in the following empirical analysis, the American
Physical Society (APS) and the Web of Science (WoS) dataset
(SI Appendix, section~S1). The APS journals are the major publication outlets
in physics. WoS includes papers in both sciences and social sciences. We focus
on the $384,649$ papers in the APS and $22,379,244$ papers in the WoS that
received at least one citation. Those papers span more than a century, and thus
allow us to investigate the SB phenomenon for a long observation period. Whereas
the APS dataset can be viewed as a perfect proxy to characterize citation
dynamics within the monodisciplinary research field of physics and is used to
compare our analysis with a previous study~\cite{Redner-110citation-2005}, the
WoS dataset allows us to underpin multidisciplinary features of the SB phenomenon.

\section{Results}

\subsection{Sleeping Beauties in physics}

\begin{figure}
\begin{center}
\includegraphics[trim=0mm 0mm 0mm 0mm, width=\columnwidth]{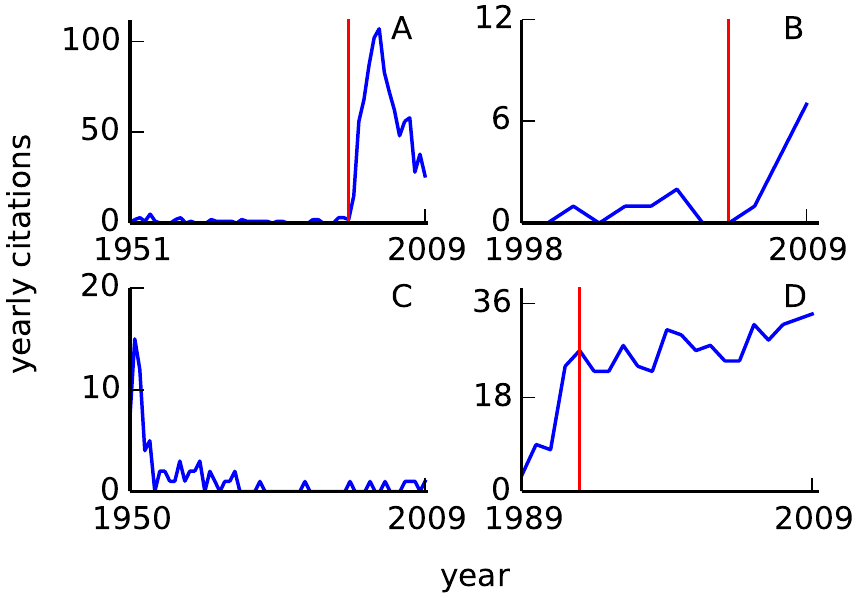}
\caption{\label{fig:beauty-diff}Dependence of the beauty coefficient on
citation history. Blue curves show yearly citations of four papers with
different $B$ values in the American Physical Society (APS) dataset:
(\emph{A}) Phys Rev \textbf{82}:403 (1951), $B = 1,722$~\cite{PhysRev.82.403};  
(\emph{B}) Phys Rev B \textbf{58}:12547 (1998), $B = 22$~\cite{PhysRevB.58.12547};
(\emph{C}) Phys Rev \textbf{78}:294 (1950), $B = 0$~\cite{PhysRev.78.294};
(\emph{D}) Phys Rev Lett \textbf{62}(3):324 (1989), $B = -5$~\cite{PhysRevLett.62.324}.
Red lines indicate their awakening time. The awakening year in \emph{C} is
$1950$, i.e., $t_a = 0$.
}
\end{center}
\end{figure}

First, we qualitatively demonstrate the resolution
power of $B$ for four papers
with radically different citation trajectories.
Fig.~\ref{fig:beauty-diff}\emph{A} shows
a paper with a very high $B$ value. Published in 1951, this paper
collected a small number of yearly citations
until $1994$, when it suddenly started to receive many citations until reaching
its maximum in $2000$.
Fig.~\ref{fig:beauty-diff}\emph{B} exhibits a
qualitatively similar citation
trajectory for a recently published paper with a very low $c_{t_m}$ and
consequently a much smaller $B$.
The paper in Fig.~\ref{fig:beauty-diff}\emph{C}
achieved its maximum yearly citations at $t = 1$.
The citation history $c_t$
therefore coincides with the reference line $\ell_t$ in $0 \leq t \leq t_m$,
yielding $B = 0$. Note that our measure $B$ only examines how the citation
curve reaches its peak, but does not consider how it decreases after that.
The paper in Fig.~\ref{fig:beauty-diff}\emph{D} is characterized by a negative $B$
value, as $c_t$ is above the reference line.

Second, we test the effectiveness of $B$ to identify top SBs in the APS by using the $12$ revived classics, previously identified by Redner, as a benchmark set~\cite{Redner-110citation-2005}.
Our results are in excellent agreement with Redner's analysis~\cite{Redner-110citation-2005}: 6 out 12 of the revived classics detected by Redner are in our top 10 list; the other 6 have also very high $B$ values, although they occupy less important positions in the ranking according to $B$ (SI Appendix, Table~S1).
Differences are due to the principles underlying the two approaches, with ours not relying on threshold parameters for the sleeping time and the number of citations.
To better clarify the diversity of the two approaches, SI Appendix, Figs.~S2 and~S3 report the citation history of the $24$ papers with highest $B$ values in the APS dataset.
We see that our measure identifies papers with a long hibernation period followed by a sudden burst in yearly citations,
without the need to reach extremely high values of citations.
As already pointed out by Redner~\cite{Redner-110citation-2005}, the list of top SBs in the APS reveals a natural grouping into a relatively small number of coarse topics, with papers belonging to the same topic exhibiting remarkably similar citation histories (SI Appendix, Fig.~S11). 
This suggests that a ``premature'' topic may fail to attract the community attention even when it is introduced by authors who have already established a strong scientific reputation. 
A corroborating evidence is provided by the famous EPR paradox paper
by Einstein, Podolsky, and Rosen that is among the top SBs we found in this dataset (SI Appendix, Fig.~S2\emph{B}).

\subsection{How rare are Sleeping Beauties?}

In contrast with previous SB definitions~\cite{Glanzel-beauty-03,
Raan-Beauty-04, Redner-110citation-2005}, ours does not rely on the arbitrary
choice of age or citation thresholds. This fact puts us in the unique position
of investigating the SB phenomenon at the systemic level and asking fundamental
questions from the macroscopic point of view: Are papers with extreme values
of $B$ exceptional occurrences? Do the majority of papers behave in a
qualitatively different way from the extreme cases discussed above, when their
sleeping period and bursty awakening are considered?
 
\begin{figure}
\begin{center}
\includegraphics[trim=0mm 0mm 0mm 0mm, width=\columnwidth]{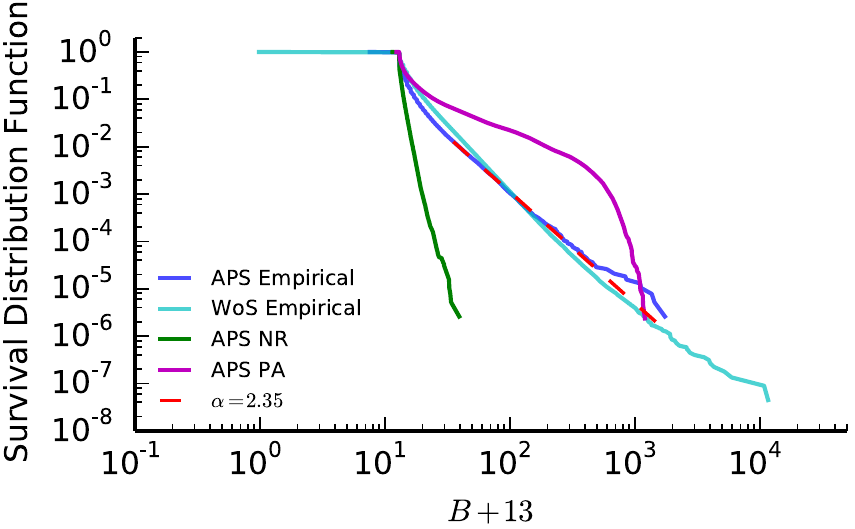}
\caption{\label{fig:beauty-ccdf}Survival distribution functions of beauty
coefficients. On the horizontal axis, we shift the values by $13$ (i.e., the
minimal value of $B$ is $-12.02$) to make all points visible in the logarithmic
scale. The blue and cyan curves represent the empirical results obtained on the
APS and WoS datasets, respectively. Results obtained with the NR and PA model
are plotted as green and magenta lines, respectively. The red dashed line
stands for the best estimate of a power-law fit of the APS curve: exponent
$\alpha = 2.35$ and the minimum value of the range of the fit $B_m = 22.27$ are
estimated using the statistical methods developed by Clauset
et~al.~\cite{Clauset-powerlaw-09}. In the APS and WoS, $4.68\%$ and $6.56\%$ of
papers, respectively, have negative $B$ values.}
\end{center}
\end{figure}

To this end, we provide a statistical description of the distribution of beauty
coefficients across all papers in each of the two datasets. Fig.~\ref{fig:beauty-ccdf}
shows the survival distribution functions of $B$ for
all papers in the APS and WoS datasets. We observe a heterogeneous but
continuous distribution of $B$, spanning several orders of magnitude. 
Except for the cutoff---which is much larger for the WoS dataset---APS and WoS exhibit remarkably similar distributions.
Although
the vast majority of papers exhibit low values of $B$, there is a consistent
number of papers with high $B$. The distributions also show no typical value or
mode; there are no clear demarcation values that allow us to separate SBs from
``normal'' papers: delayed recognition occurs on a wide and continuous range,
in sharp contrast with previous results claiming that SBs are extraordinary
cases~\cite{Glanzel-beauty-03, Glanzel-beauty-04, Redner-110citation-2005}.

It may appear as not entirely fair to compare beauty
coefficients for papers of different ages~\cite{marx2010reference}:
Later papers have by definition less
chance to develop a long sleeping period and to exhibit a sudden awakening.
This may, to some extent, dictate the shapes of observed distributions. On the
other hand, the vast majority of papers tend to have a single and well-defined
peak in their yearly citations early during their lifetime, implying that their
$B$ values do not change with moving the observation time $T$ far into the
future. In particular, our estimations indicate that
nearly $90\%$ of the papers
have already experienced a drastic decrement
after their maximum number of yearly citations,
irrespective of their $B$ value (SI Appendix, section~S3). The shapes of the empirical distributions remain essentially unchanged 
if we consider only the papers that have experienced the typical sharp decline of the post-maximum yearly citation rate.

\subsection{Is the Sleeping Beauty phenomenon statistically significant?}

The result of the previous section implicitly suggests that the SB phenomenon
could be in principle described via a simple mechanism that works essentially
at all scales. This leads naturally to the question whether the observed
distributions of $B$ can be accounted for by idealized network evolution
models. To address this question, we first consider a citation network
randomization (NR) process where citations are randomly reshuffled, preserving
time order (SI Appendix, section~S4). SI Appendix, Fig.~S2 compares the citation history of the
top nine SBs in the APS dataset and the corresponding ones obtained through the
NR process. They typically show opposite trends, with NR histories exhibiting a
rapid decline. This is not surprising: As later papers are considered, the
probability for an existing paper to receive a citation from one of such late
papers decreases simply because there is a larger number of papers that could
potentially receive the citation. This leads to typically smaller beauty
coefficients, as evident in the sharp decrease of the NR distribution in
Fig.~\ref{fig:beauty-ccdf}, and the associated small maximum value $B=30$.

Next, we consider the preferential attachment (PA) mechanism as another
baseline model, as it is one of the most fundamental ingredients used in most
modeling efforts aimed at describing citation histories of papers. In the PA
baseline, references of progressively added citing papers are reassigned
according to the PA mechanism (SI Appendix, section~S4). SI Appendix, Fig.~S2 also shows slowly
increasing yearly citations by the PA model, explained by the positive feedback
effect generated via the PA mechanism. The overall number of citations
according to PA baseline for the nine papers in SI Appendix, Fig.~S2 remains small. Those
are relatively young papers in the dataset and their probability to receive
citations, according to PA, is reduced by that of older papers. The resulting
distribution of $B$ in Fig.~\ref{fig:beauty-ccdf} shows a much smaller range
and a well-defined cutoff. It remains to be seen to what extent a recently
proposed model for citation histories~\cite{Wang-impact-13} are compatible with
the SB phenomenon.

\subsection{Sleeping Beauties in science}

The occurrence of extreme cases of SBs is not limited to physics.
Table~\ref{tab:beauty-wos} lists basic information about the $15$ papers with
the highest $B$ values in the WoS dataset (see SI Appendix, Fig.~S4 for their citation
histories). This list contains four SBs that were published in the 1900s.
Consistent with previous studies, we find that many SBs are in the field of
physics and chemistry~\cite{Glanzel-beauty-03}. Two papers are, however, in the
field of statistics, which fails to be noted before as a top discipline
producing SBs. One of them slept for more than one century: the paper by the
influential statistician Karl Pearson, published in $1901$ in the journal
\emph{Philosophical Magazine}, shows the relation between principal component
analysis and the minimization chi-distance. 
The other one, published in $1927$ (therefore sleeping
for more than $70$ years), introduces the Wilson score interval, one type of
confidence interval for estimating a proportion that improves over the commonly
used normal approximation interval.
The 3rd ($B=5,923$),
12th ($B=2,584$),
and 15th ($B=2,184$)
top-ranked papers
in the WoS dataset were published in \emph{Physical Review}, but were not
ranked as top papers in the APS dataset, suggesting that the bulk of their
citations are mainly from journals not contained in the APS dataset. The EPR
paradox paper (the 14th),
however, is ranked at the top in both datasets.

\begingroup
\squeezetable
\begin{table*}
\centering
\caption{\label{tab:beauty-wos}Top 15 SBs in science.
From left to right, we report for each paper its beauty coefficient $B$,
author(s) and title, publication and awakening year, publication journal,
and scientific domain. See SI Appendix, Fig.~S4 for detailed citation histories of these
papers.}
\begin{tabular*}{\hsize}{@{\extracolsep{\fill}}c l l c l c}
\hline
$B$ & Author(s) & Title & Pub., awake & Journal & Field\cr \hline
11600 & Freundlich, H & Concerning adsorption in solutions & 1906, 2002 & Z. Phys. Chem. & Chem.\cr \hline
\multirow{2}{*}{10769} & Hummers, WS & \multirow{2}{*}{Preparation of Graphitic Oxide} & \multirow{2}{*}{1958, 2007} & \multirow{2}{*}{J. Am. Chem. Soc.} & \multirow{2}{*}{Chem.}\cr
& Offeman, RE & & & &\cr \hline
\multirow{2}{*}{5923} & \multirow{2}{*}{Patterson, AL} & The Scherrer formula for x-ray particle size & \multirow{2}{*}{1939, 2004} & \multirow{2}{*}{Phys. Rev.} & \multirow{2}{*}{Phys.}\cr
& & determination & & &\cr \hline
\multirow{2}{*}{5168} & Cassie, ABD & \multirow{2}{*}{Wettability of porous surfaces} & \multirow{2}{*}{1944, 2002} & \multirow{2}{*}{Trans. Faraday Soc.} & \multirow{2}{*}{Chem.}\cr
& Baxter, S & & & &\cr \hline
\multirow{3}{*}{4273} & Turkevich, J & A study of the nucleation and growth & \multirow{3}{*}{1951, 1997} & \multirow{3}{*}{Discuss. Faraday Soc.} & \multirow{3}{*}{Chem.}\cr
& Stevenson, PC & processes in the synthesis of colloidal & & &\cr
& Hillier, J    & gold & & &\cr \hline
\multirow{2}{*}{3978} & \multirow{2}{*}{Pearson, K} & On lines and planes of closest fit to & \multirow{2}{*}{1901, 2002} & \multirow{2}{*}{Philos. Mag.} & \multirow{2}{*}{Statist.}\cr
& & systems of points in space & & &\cr \hline
3892 & Stoney, GG & The tension of metallic films deposited by electrolysis & 1909, 1989 & Proc. R. Soc. Lond. A & Phys.\cr \hline
3560 & Pickering, SU & CXCVI.--Emulsions & 1907, 1998 & J. Chem. Soc., Trans. & Chem.\cr \hline
2962 & Wenzel, RN & Resistance of solid surfaces to wetting by water & 1936, 2003 & Ind. Eng. Chem. & Chem.\cr \hline
\multirow{2}{*}{2736} & \multirow{2}{*}{Wilson, EB} & Probable inference, the law of succession, & \multirow{2}{*}{1927, 1999} & \multirow{2}{*}{J. Am. Statist. Assoc.} & \multirow{2}{*}{Statist.}\cr
& & and statistical inference & & &\cr \hline
\multirow{2}{*}{2671} & \multirow{2}{*}{Langmuir, I} & The constitution and fundamental properties & \multirow{2}{*}{1916, 2003} & \multirow{2}{*}{J. Am. Chem. Soc.} & \multirow{2}{*}{Chem.}\cr
& & of solids and liquids Part I Solids & & &\cr \hline
\multirow{2}{*}{2584}  & Moller, C; & Note on an approximation treatment for & \multirow{2}{*}{1934, 1982} & \multirow{2}{*}{Phys. Rev.} & \multirow{2}{*}{Phys.}\cr
& Plesset, MS & many-electron systems & & &\cr \hline
\multirow{2}{*}{2573} & \multirow{2}{*}{Pugh, SF} & Relations between the elastic moduli and the & \multirow{2}{*}{1954, 2005} & \multirow{2}{*}{Philos. Mag.} & \multirow{2}{*}{Metallurgy}\cr
& & plastic properties of polycrystalline pure metals & & &\cr \hline
\multirow{3}{*}{2258} & Einstein, A & Can quantum-mechanical description of & \multirow{3}{*}{1935, 1994} & \multirow{3}{*}{Phys. Rev.} & \multirow{3}{*}{Phys.}\cr
& Podolsky, B & physical reality be considered complete? & & &\cr
& Rosen, N & & & &\cr \hline
2184  & Washburn, EW & The dynamics of capillary flow & 1921, 1995 & Phys. Rev. & Phys.\cr
\hline
\end{tabular*}
\end{table*}
\endgroup

SI Appendix, Tables~S2 and S3 list basic information about the top $10$ SB papers in
statistics and mathematics, respectively. Publications introducing many
important techniques, like Fisher's exact test, Metropolis--Hastings
algorithm, and Kendall rank correlation coefficient, have high beauty
coefficients. We also find numerous examples of SBs in the social sciences
(SI Appendix, Table~S4), in contrast with previous results about their
alleged absence~\cite{Glanzel-beauty-03}.

\begin{figure}
\begin{center}
\includegraphics[trim=0mm 0mm 0mm 0mm, width=\columnwidth]{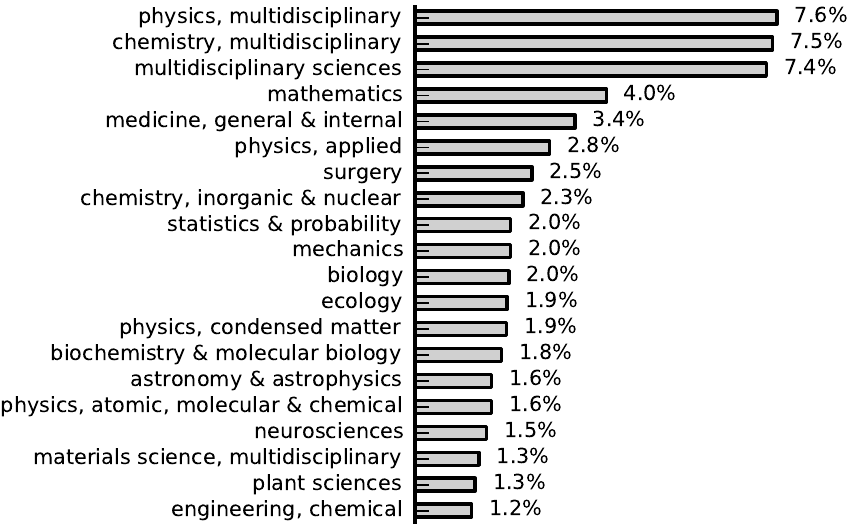}
\caption{\label{fig:top-beauty-cat}Top 20 disciplines producing SBs in science.
We consider papers with beauty coefficient in the top $0.1\%$ of the entire WoS
database, and compute the fraction of those papers that fall in a given
subject category.}
\end{center}
\end{figure}

How are SBs distributed among different (sub-)disciplines? To further investigate the
multidisciplinary character of the
SB phenomenon, we took advantage of journal
classifications provided by Journal Citation Reports (JCR)
(thomsonreuters.com/en/products-services/scholarly-scientific-research/research-managementand-evaluation/journal-citation-reports.html),
which
classify scientific journals into one or more subject categories (e.g.
physics, multidisciplinary; mathematics; medicine, general and internal). We
first consider only papers published in journals belonging
to at least one JCR subject category, and focus on the top $0.1\%$ of papers
with highest $B$ values. Then, we compute the fraction of those papers that
belong to a given subject category. Fig.~\ref{fig:top-beauty-cat} shows the top
$20$ categories producing SBs. Subfields of physics, chemistry, and mathematics are
noticeably the top disciplines, consistently with previous
studies~\cite{Glanzel-beauty-03}. Some disciplines not previously noted include
medicine (internal and surgery), statistics and probability. Particularly
interesting is the category multidisciplinary sciences, ranked third,
that includes top journals like \emph{Nature}, \emph{Science}, and \emph{PNAS},
because (\emph{i}) delayed recognition signals that such contributions may be
perceived by the academic community as too premature or futuristic, although it
is common ground among academics to speculate that such venues only publish
trending topics, and (\emph{ii}) journals in the multidisciplinary sciences subject
category are really more fit to attract publications that become field-defining
even decades after their appearance.

\subsection{What triggers the awakening of an SB?}

A full answer to this question would require a case-by-case examination, but it can be addressed in a systematic way by studying the papers that cite the SB before and after its awakening. 
To illustrate this strategy, it is worth to examine two paradigmatic examples of top SBs.

\begin{figure}
\begin{center}
\includegraphics[trim=0mm 0mm 0mm 0mm, width=\columnwidth]{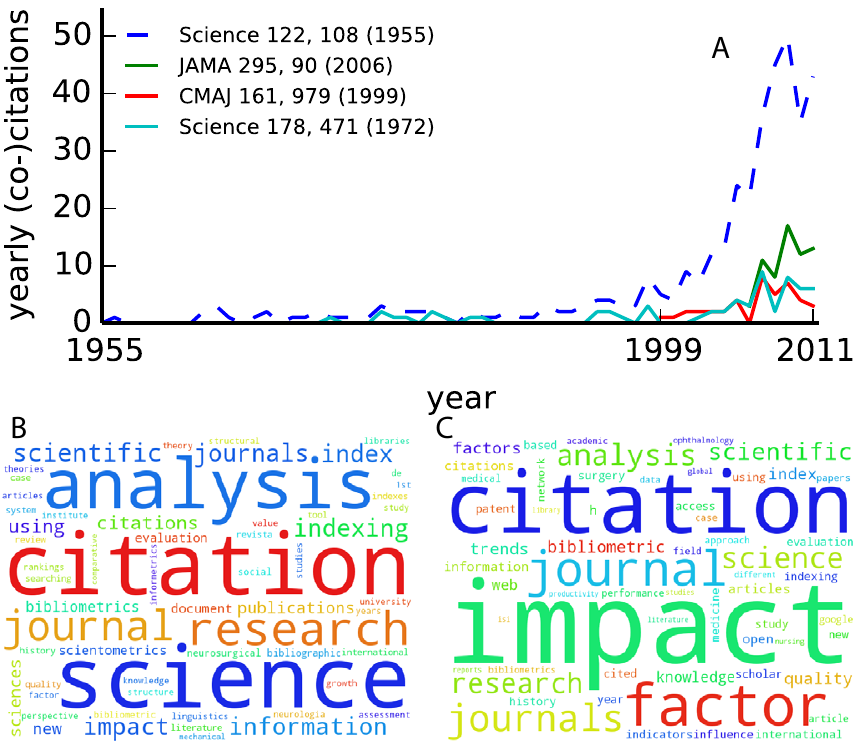}
\caption{\label{fig:garfield}Paradigmatic example of the awakening of an SB.
(\emph{A}, blue) Citation history of the paper Science \textbf{122}:108 (1955)~\cite{Garfield-sci-1955}.
The three most co-cited papers are
green, JAMA \textbf{295}:90 (2006)~\cite{Garfield-jama-2006};
cyan, Science \textbf{178}:471 (1972)~\cite{Garfield-science-1972}; and
red, Can Med Assoc J \textbf{161}:979 (1999)~\cite{Garfield-cmaj-1999}.
(\emph{B} and \emph{C}) Clouds of the most frequent keywords appearing in the title of papers citing
Science \textbf{122}:108 (1955)~\cite{Garfield-sci-1955}, published,
respectively, before (\emph{B}) and after (\emph{C}) year $2000$.}
\end{center}
\end{figure}

The first is the 1955 Garfield paper introducing the ancestor of the Web of Science database~\cite{Garfield-sci-1955}.
This paper slept for almost $50$ years, becoming suddenly popular around $2000$. 
A simple investigation based on co-citations, similar to the one performed in ref.~\cite{marx2014shockley}, reveals that the delayed recognition of the 1955 paper by Garfield was triggered by later articles by the same author (Fig.~\ref{fig:garfield}\emph{A}). 
Such papers, in turn, were cited by very influential works in two different contexts: (\emph{i}) the 1999 article by Kleinberg about the hyperlink-induced topic search (HITS) algorithm, which can be considered one pioneering works in network science~\cite{kleinberg1999authoritative}; and (\emph{ii}) the 1998 paper by Seglen on the limitations of the journal impact factor, which historically represents the beginning of the ongoing debate
about the (mis)use of citation indicators in research evaluation~\cite{seglen1997impact}. 
The change in contextual importance of the 1955 paper by Garfield is further revealed by the frequency of keywords appearing in the titles of its citing papers before and after year 2000 (Fig.~\ref{fig:garfield}\emph{B} and \emph{C}), with the notion of ``impact factor'' becoming the main recognizable difference.
With a similar motivation, the 1977 paper by Zachary also tops the ranking of SBs coming from the social sciences~\cite{Zachary-karate-1977}.
This paper was essentially unnoticed for about $30$ years, but then became suddenly important in network science research after the publication of the seminal paper by Girvan and Newman, which adopts the social network described in the Zachary paper as a paradigmatic benchmark to validate community detection methods on graphs~\cite{Girvan-comm-2002} (SI Appendix, Fig.~S12).

\begin{figure}
\begin{center}
\includegraphics[trim=0mm 0mm 0mm 0mm, width=\columnwidth]{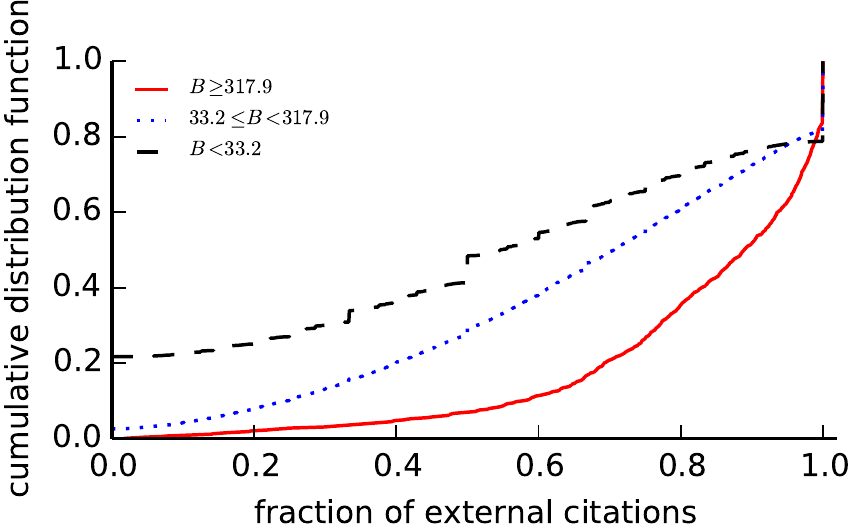}
\caption{\label{fig:interdisciplinary}Interdisciplinary nature of top SBs.
Cumulative distribution functions of fraction of external citations for the group of (red) top
$1,000$ SBs ($B \geq 317.93$); (blue) from the 1,001st to the top $1\%$
($33.21 \leq B < 317.93$); and (black) the rest ($B < 33.21$). The horizontal axis measures for
each paper the fraction of its citations that originate from other subject
categories.}
\end{center}
\end{figure}

The examples above suggest that a partial explanation behind the sudden awakening of top SBs may lie in the fact that the paper in question is suddenly ``discovered'' as relevant by an entire community in another discipline.
To support this hypothesis, in Fig~\ref{fig:interdisciplinary} we divide the papers in the WoS dataset in three disjoint subsets with high, medium, and low values of $B$.
For each subset we compute the cumulative distribution for the fraction of citations received by a paper from publications in a discipline (as inferred by the journal of publication) different from that of the cited paper.
Top SBs are clearly different from the other two categories and are characterized by a typically very high fraction of citations from other disciplines: for about $80\%$ of the top SBs, as much as $75\%$ or more of citations are of interdisciplinary nature.

\section{Discussion}

The main purpose of this work was to introduce a parameter-free method to
quantify to what extent a paper is an SB. Through a systematic
analysis carried out on large-scale bibliographic databases and over
observation windows longer than a century, we have shown that our method
correctly identifies cases that meet the intuitive notion of SBs. We noticed
that our measure is not entirely free of biases: Comparing the degree of
beauty between papers in different disciplines or ages may be
problematic due to differences in the overall citation patterns. Despite this
limitation, we found that papers whose citation histories are characterized by
long dormant periods followed by fast growths are not exceptional outliers, but
simply the extreme cases in very heterogeneous but otherwise continuous
distributions. Simple models based on cumulative advantage, although consistent
with overall citation distributions, are not easily reconciled with the
observed distributions of beauty coefficients. Further work is needed to
uncover the general mechanisms that may be held responsible for the awakening
of SBs.

\begin{acknowledgments}
We thank Claudio Castellano, Filippo Menczer, Yong-Yeol Ahn, Cassidy Sugimoto,
and Chaoqun Ni for insightful discussions, and the American Physical Society
for making the APS dataset publicly available. This work is partially supported
by NSF (grant SMA-1446078).
\end{acknowledgments}

\clearpage

\begin{center}\LARGE\textbf{Supporting Information}\end{center}

\setcounter{figure}{0}
\makeatletter 
\renewcommand{\thefigure}{S\@arabic\c@figure}
\makeatother

\setcounter{table}{0}
\makeatletter 
\renewcommand{\thetable}{S\@arabic\c@table}
\makeatother

\setcounter{section}{0}
\makeatletter 
\renewcommand{\thesection}{S\@arabic\c@section}
\makeatother

\section{Datasets}

\begin{figure}
\begin{center}
\includegraphics[trim=0mm 0mm 0mm 0mm, width=0.9\columnwidth]{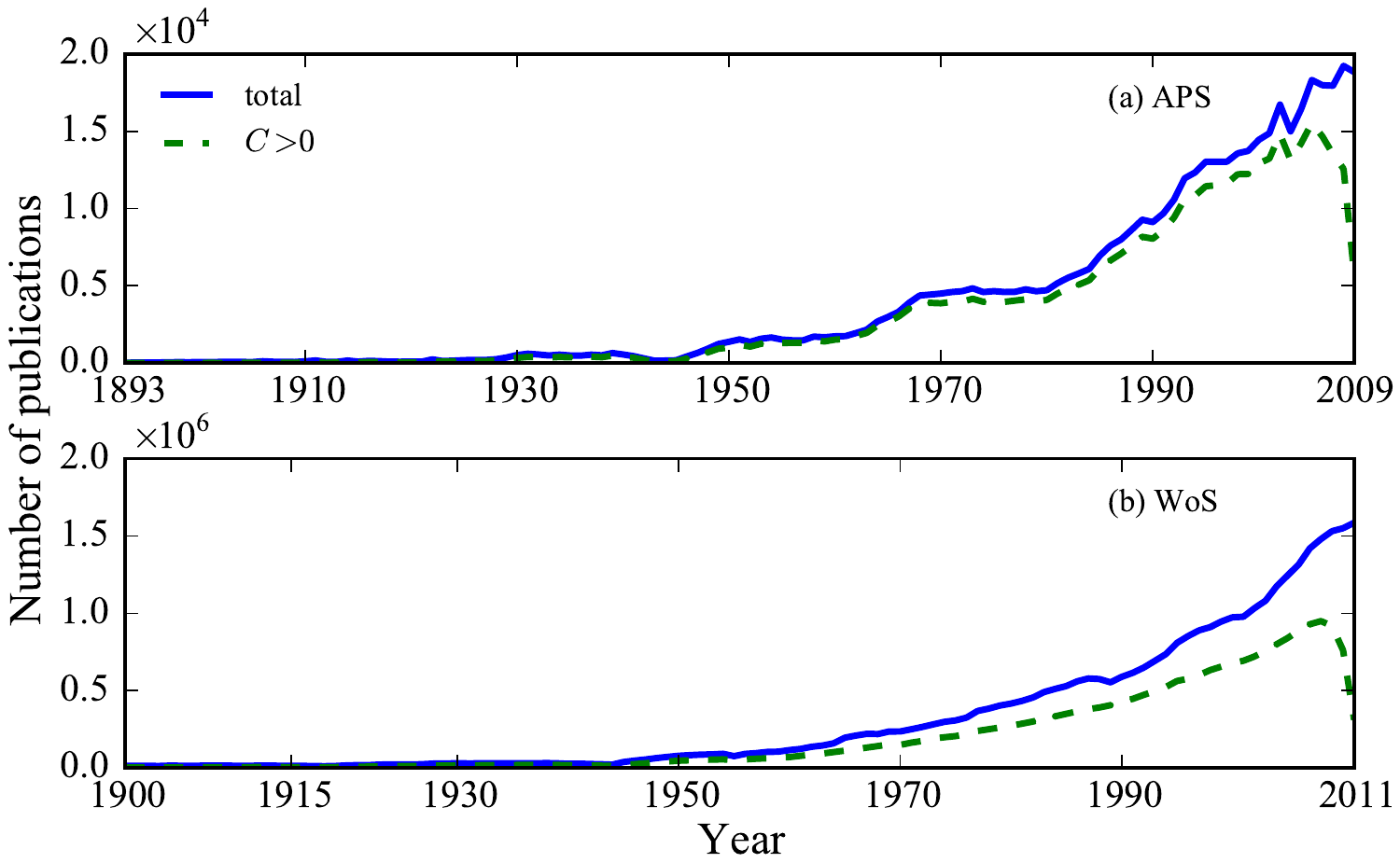}
\caption{\label{fig:pub-per-year}(Blue solid) Total number of papers per year;
(Green dashed) Yearly number of papers that received citations.}
\end{center}
\end{figure}

In this work, we use two large datasets, namely the American Physical Society
(APS) and the Web of Science (WoS). APS contains $463,348$ papers published
from $1893$ to $2009$ in APS journals and is publicly available upon request
at \url{http://journals.aps.org/datasets}; WoS is comprised of $35,174,034$
papers published between $1900$ and $2011$ in journals covering most research
fields, and is available upon purchase from Thomson Reuters. Most papers in the
APS dataset are also in the WoS. The APS dataset, though, contains fewer
citations: only those originating from papers within the APS journals are
therein recorded. Our analysis is based on papers that received at least one
citation. A total number of $384,649$ and $22,379,244$ such papers were found
in the APS and WoS dataset, respectively. Fig.~\ref{fig:pub-per-year} shows the
yearly number of papers with at least one citation received before the end of
the observation period. The fact that recent papers have had less time to
accumulate citations is reflected in the sharp decrease that is noticeable as
time approaches the end of the observation period.

\section{Examples of top Sleeping Beauties}

\begin{figure*}
\begin{center}
\includegraphics[trim=0mm 0mm 0mm 0mm, width=0.9\textwidth]{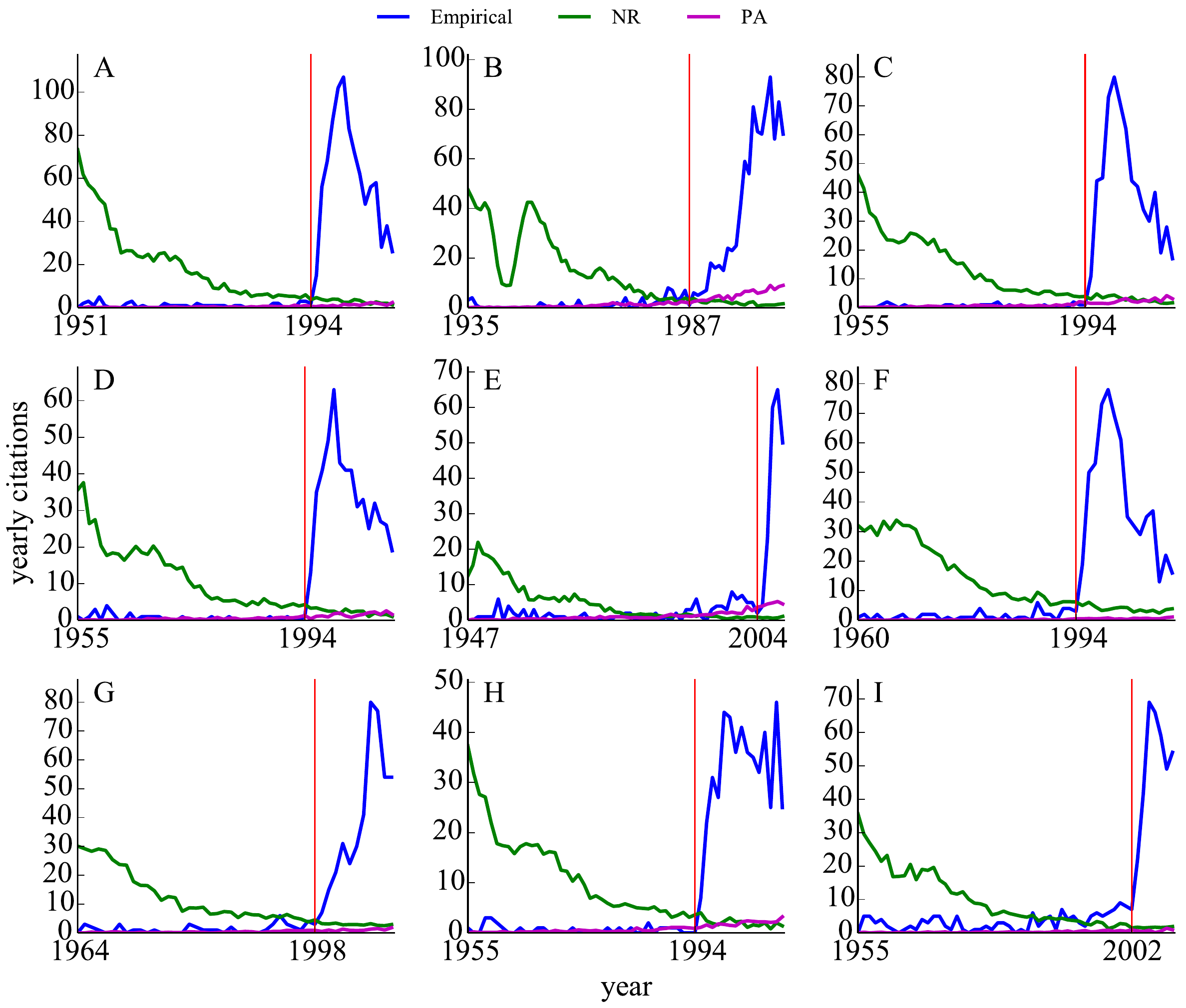}
\caption{\label{fig:beauty-top}Top Sleeping Beauties in physics.
Blue curves show yearly citations received by papers:
(\emph{A}) Phys. Rev. 82, 403 (1951), $B = 1,722$~\cite{PhysRev.82.403-1};
(\emph{B}) Phys. Rev. 47, 777 (1935), $B = 1,419$~\cite{PhysRev.47.777};
(\emph{C}) Phys. Rev. 100, 675 (1955), $B = 1,348$~\cite{PhysRev.100.675};
(\emph{D}) Phys. Rev. 100, 545 (1955), $B = 1,107$~\cite{PhysRev.100.545};
(\emph{E}) Phys. Rev. 71, 622 (1947), $B = 1,086$~\cite{PhysRev.71.622};
(\emph{F}) Phys. Rev. 118, 141 (1960), $B = 841$~\cite{PhysRev.118.141};
(\emph{G}) Phys. Rev. 135, A550 (1964), $B = 825$~\cite{PhysRev.135.A550};
(\emph{H}) Phys. Rev. 100, 564 (1955), $B = 670$~\cite{PhysRev.100.564};
(\emph{I}) Phys. Rev. 100, 580 (1955), $B = 624$~\cite{PhysRev.100.580}.
Yearly citations obtained from citation network randomization (NR) and
preferential attachment (PA) model are plotted as green and purple lines,
respectively. Both the NR and PA results are averaged across 10 realizations.
The awakening years, identified using Eq.~3, are indicated by
the vertical red lines. The sharp decrease of the curve for the NR result in
panel \emph{B} is probably due to the decrease of number of publications during the
period of World War II (Fig.~S1a). Panels \emph{A}, \emph{C}, \emph{D}, \emph{F}, and \emph{H} refer to papers
about the double exchange mechanism. Panel \emph{B} refers to the EPR paradox paper by
Einstein, Podolsky, and Rosen. Panel \emph{E} considers the pioneering study on the band
structure of graphite.}
\end{center}
\end{figure*}

\begin{figure*}
\begin{center}
\includegraphics[trim=0mm 0mm 0mm 0mm, width=0.9\textwidth]{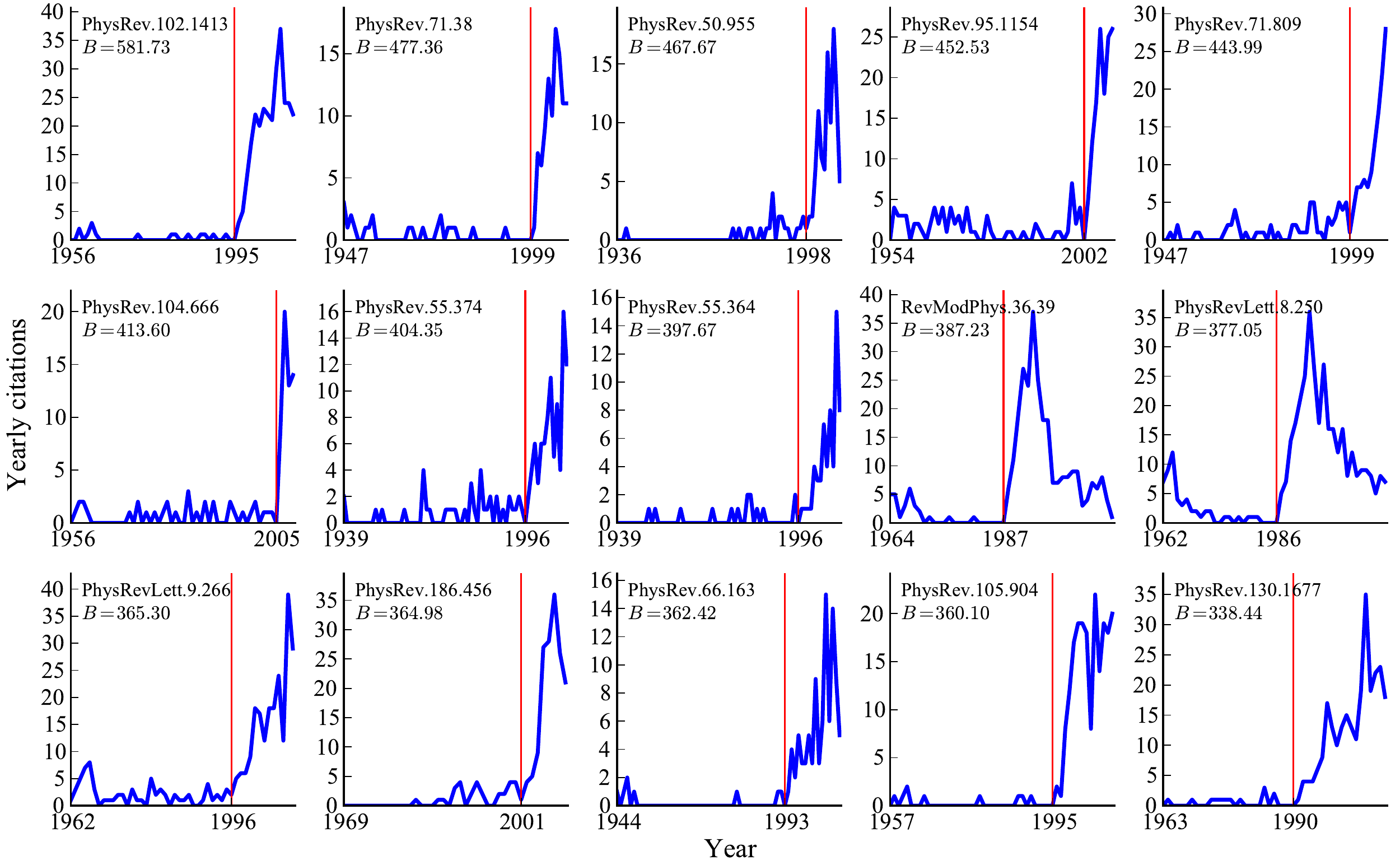}
\caption{\label{fig:beauty-top-2}(Blue) Citation histories, (Red) awakening
years, and $B$ values of the 15 papers ranked from $10^{th}$ to $24^{th}$ based
on the $B$ values in the APS dataset. The ending year is 2009.}
\end{center}
\end{figure*}

Figs.~\ref{fig:beauty-top} and \ref{fig:beauty-top-2} show the citation history
of the top 24 papers in the APS dataset. Table~\ref{tab:redner} presents the
comparison between our results and Redner's results~\cite{Redner-110citation-2005-1}.

\begin{figure*}
\begin{center}
\includegraphics[trim=0mm 0mm 0mm 0mm, width=0.9\textwidth]{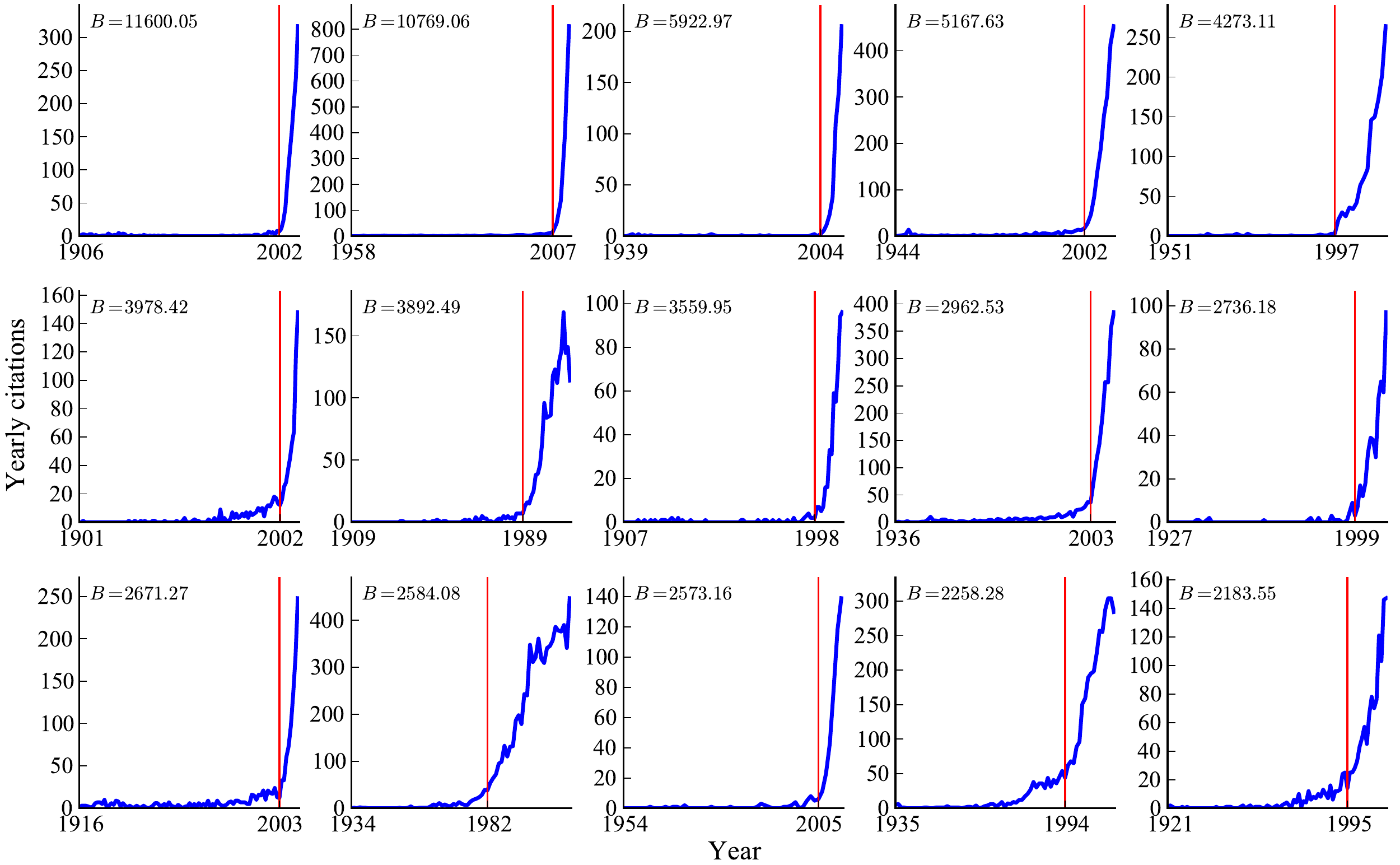}
\caption{\label{fig:beauty-wos-1}(Blue) Citation histories, (Red) awakening
years, and $B$ values of the top 15 papers, based on the $B$ values in the WoS
dataset. The ending year is 2011.}
\end{center}
\end{figure*}

\begin{figure*}
\begin{center}
\includegraphics[trim=0mm 0mm 0mm 0mm, width=0.9\textwidth]{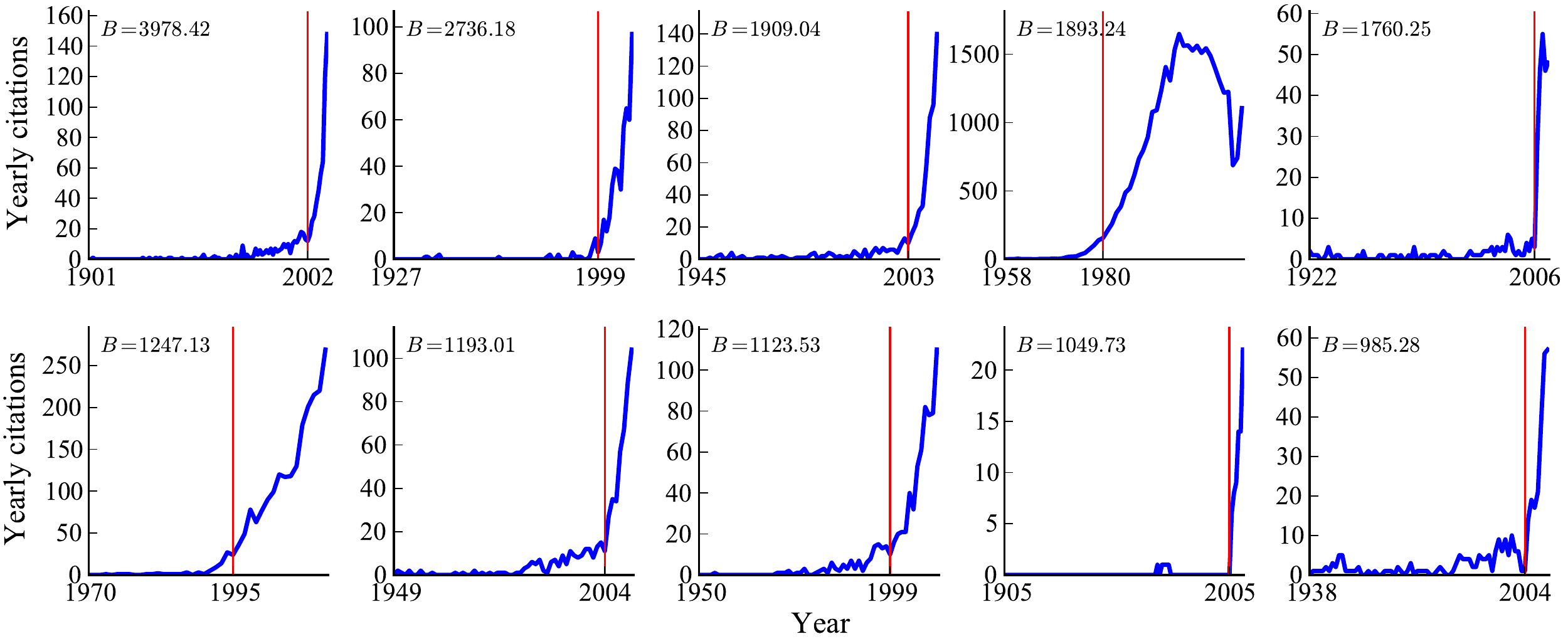}
\caption{\label{fig:beauty-wos-stats}(Blue) Citation histories, (Red) awakening
years, and $B$ values of top 10 papers in Statistics based on the $B$ values in the
WoS dataset. The ending year is 2011.}
\end{center}
\end{figure*}

\begin{figure*}
\begin{center}
\includegraphics[trim=0mm 0mm 0mm 0mm, width=0.9\textwidth]{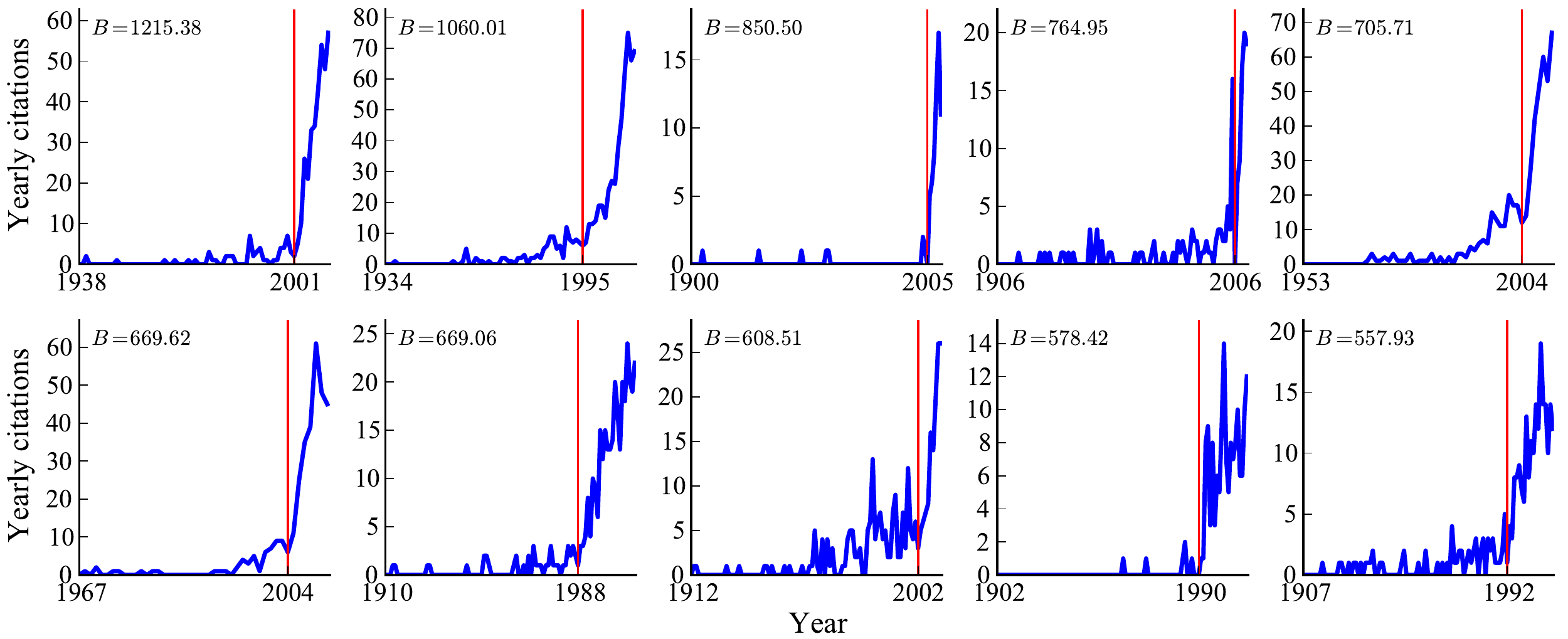}
\caption{\label{fig:beauty-wos-math}(Blue) Citation histories, (Red) awakening
years, and $B$ values of top 10 papers in Mathematics based on the $B$ values in
the WoS dataset. The ending year is 2011.}
\end{center}
\end{figure*}

\begin{figure*}
\begin{center}
\includegraphics[trim=0mm 0mm 0mm 0mm, width=0.9\textwidth]{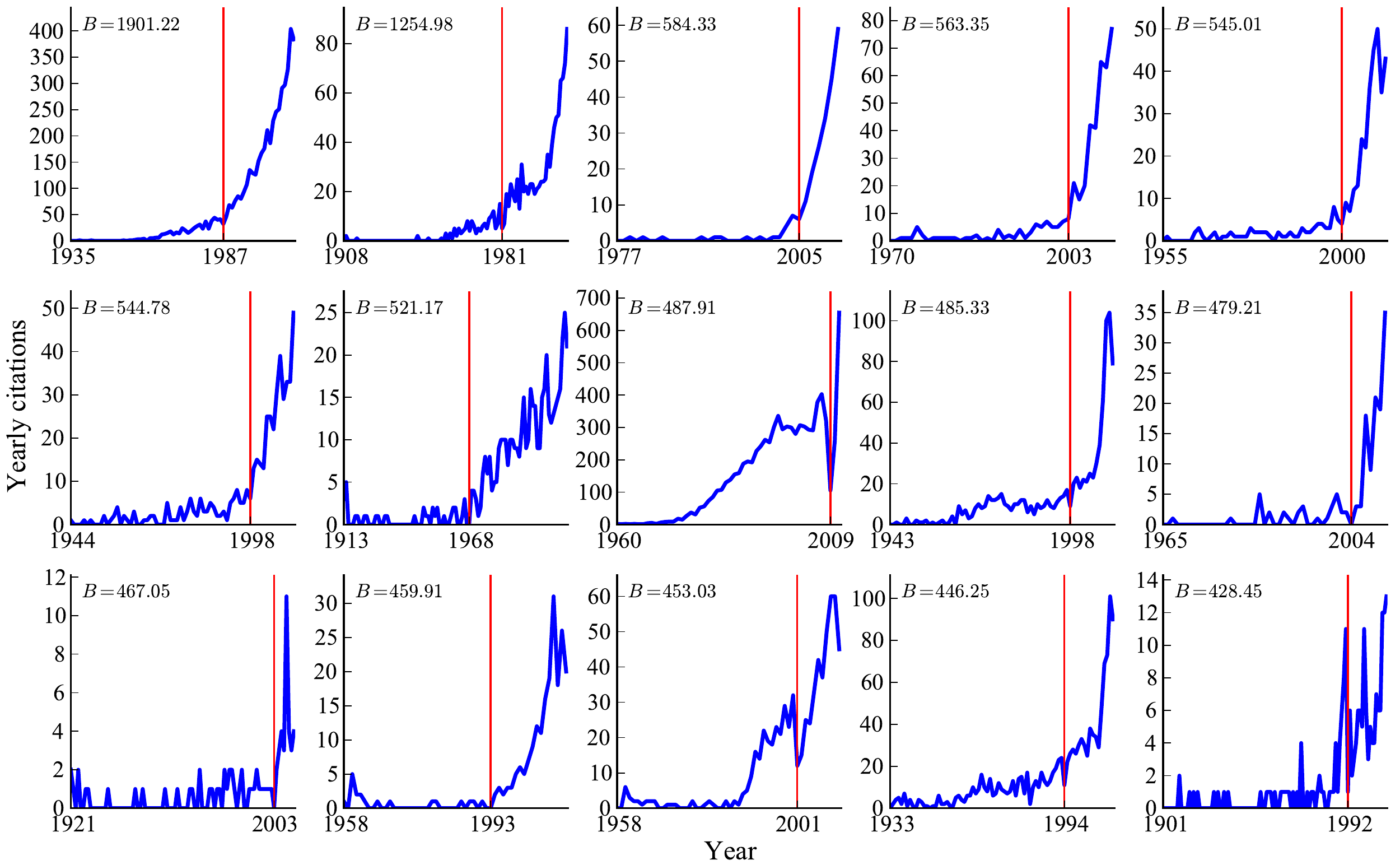}
\caption{\label{fig:beauty-wos-ssh-1}(Blue) Citation histories, (Red) awakening
years, and $B$ values of top 15 Sleeping Beauties in Social Sciences and
Humanities based on the $B$ values in the WoS dataset. The ending year is 2011.}
\end{center}
\end{figure*}

\begin{figure*}
\begin{center}
\includegraphics[trim=0mm 0mm 0mm 0mm, width=0.9\textwidth]{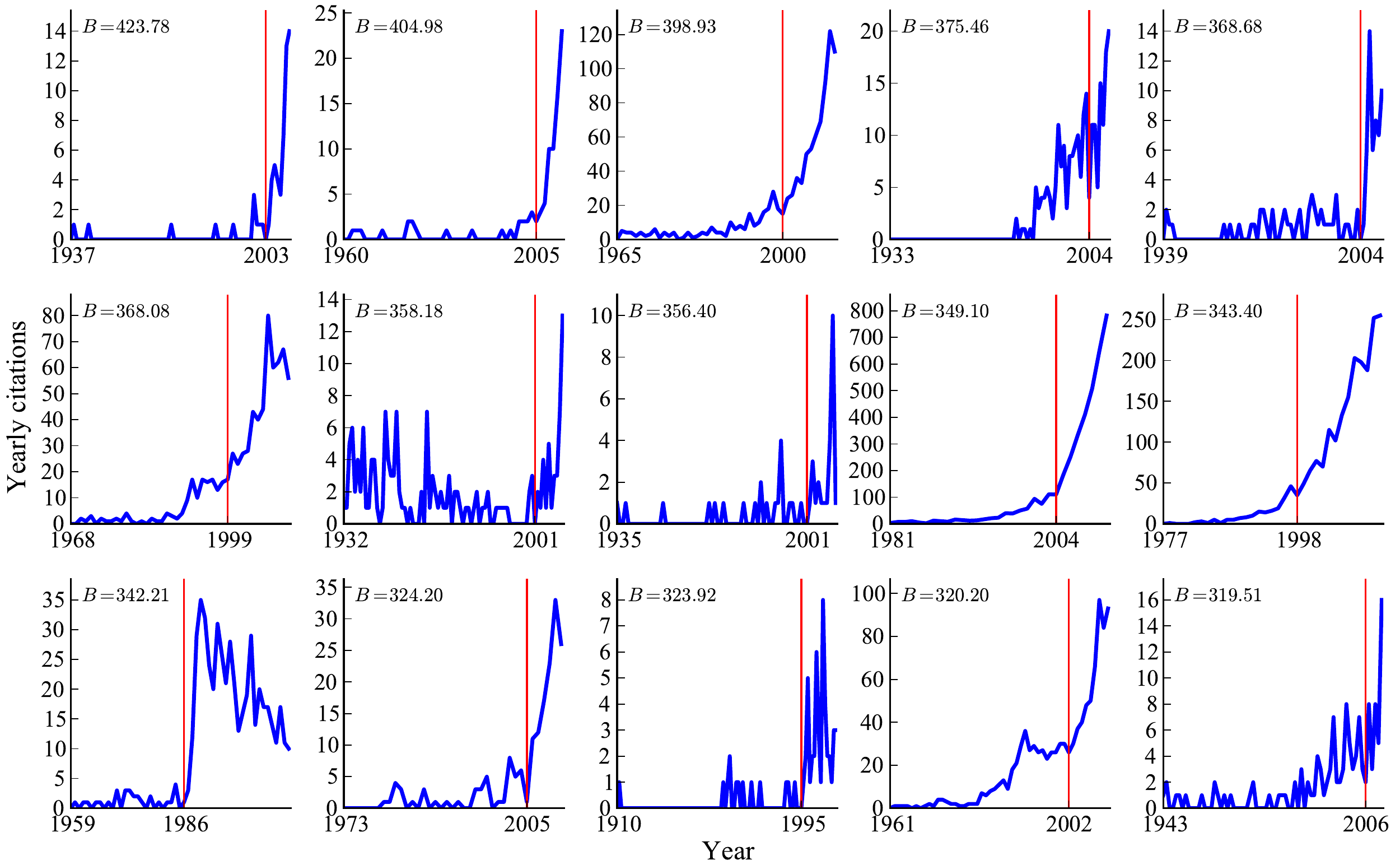}
\caption{\label{fig:beauty-wos-ssh-2}(Blue) Citation histories, (Red) awakening
years, and $B$ values of 15 Sleeping Beauties ranked from $16^{th}$ to
$30^{th}$ in Social Sciences and Humanities based on $B$ values in the WoS
dataset. The ending year is 2011.}
\end{center}
\end{figure*}

Fig.~\ref{fig:beauty-wos-1} displays the citation history of the top $15$
Sleeping Beauties in the WoS dataset showed in Table~I of the main text.
Tables~\ref{tab:beauty-wos-stats}, \ref{tab:beauty-wos-math}, and
\ref{tab:beauty-wos-ssh} present the basic information of the top Sleeping Beauties
in Statistics, Mathematics, and Social Sciences and Humanities, respectively.
See Figs.~\ref{fig:beauty-wos-stats}--\ref{fig:beauty-wos-ssh-2} for
corresponding citation histories.

\section{Characterizing Decreasing Patterns}

\begin{figure}
\begin{center}
\includegraphics[trim=0mm 6mm 0mm 0mm, width=\columnwidth]{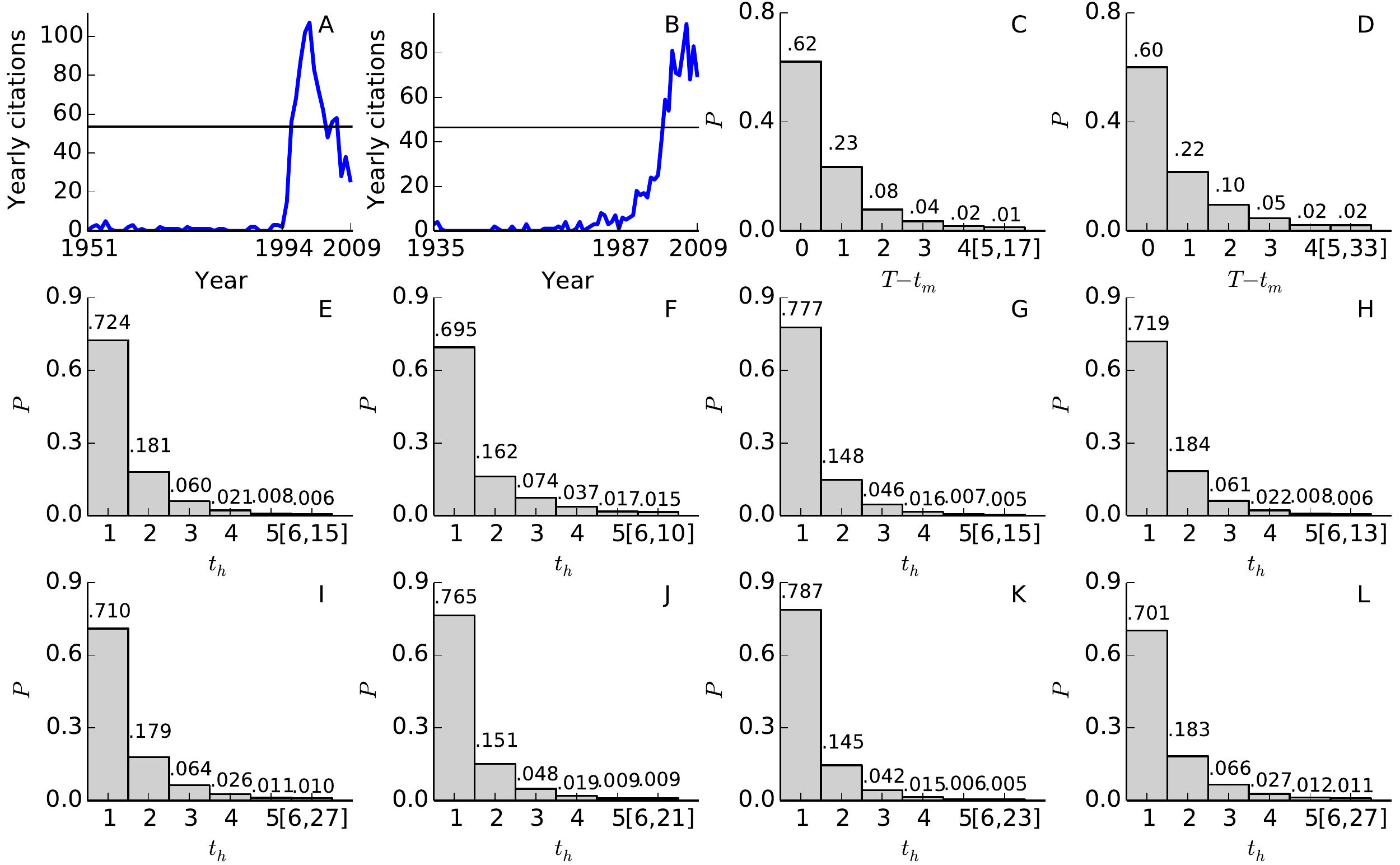}
\caption{\label{fig:beauty-half-life}Characterization of decreasing citation
patterns of Sleeping Beauties. (\emph{A--B}) Papers with positive beauty
coefficient $B$ are classified into two categories depending on whether or not
their yearly citation counts have decreased to half of their maximum.
(\emph{C--D}) For papers belonging to the first class, we measure the length
$T-t_m$ of the observation window at our disposal. $T = 2009$ for the APS and
$T = 2011$ for the WoS are the last year covered by our datasets. $t_m$ is
instead the year when we observe the maximum number of yearly citations
accumulated by an individual paper. The figures display the histograms of the
quantity $T-t_m$ obtained for the APS (\emph{C}) and WoS (\emph{D}) dataset.
(\emph{E--H}) For papers that have experienced a fall in yearly citation counts
at least below the half of their peak height $c_m$, we measure $t_h$, i.e., the
number of years necessary to fall below the line $c_m/2$. We show that the
distribution of $t_h$ is insensible to the specific dataset considered, and to
their beauty coefficient $B$. Panels \emph{F}, \emph{G} and \emph{H} refer to
the papers of the APS dataset ranked in the top $1\%$, top $1\%$ to $10\%$,
below $10\%$, respectively. Panels \emph{I--L} show the same histograms as
those of panels \emph{E--H}, but for the WoS dataset.}
\end{center}
\end{figure}

This section presents a statistical characterization of how yearly citations of
papers decrease after the peak. In summary, for most of the papers the yearly
citation rate decreases quickly (possibly exponentially) after its peak. Our
analysis focused only papers with positive beauty coefficient $B$, for a total
of $189,673$ (out of $384,649$; $49.3\%$) and $14,689,643$ (out of
$22,379,244$; $65.6\%$) papers in the APS and WoS dataset, respectively. We
further classify every of these papers into two categories depending on whether
or not their yearly citation counts $c_t$ decreased to half of its maximum
during the observation period $[t_m+1, T]$
(Figs.~\ref{fig:beauty-half-life}\emph{A-B}).

We identify $18,131$ ($9.56\%$) papers in the APS whose $c_t$ have not
decreased below $c_{t_m}/2$, and $2,094,671$ ($14.26\%$) in the WoS dataset.
Figs.~\ref{fig:beauty-half-life}\emph{C--D} display the histograms of $T-t_m$.
We observe that a large fraction are recently awakening papers, with about
$60\%$ of them getting their maximum yearly citations $c_{t_m}$ in the last
year of the observation periods ($T-t_m = 0$).

For the remaining papers whose yearly citations have decreased below
$c_{t_m}/2$, we define the paper ``half-life" $t_h$ as the number of years
required by $c_t$ to decrease from $c_{t_m}$ to $c_{t_m}/2$.
Figs.~\ref{fig:beauty-half-life}\emph{E--H} show the distributions of $t_h$
across all these papers in the APS (Fig.~\ref{fig:beauty-half-life}\emph{E}),
papers whose $B$ values ranked in the top $1\%$
(Fig.~\ref{fig:beauty-half-life}\emph{F}), from $1\%$ to $10\%$
(Fig.~\ref{fig:beauty-half-life}\emph{G}), and the rest
(Fig.~\ref{fig:beauty-half-life}\emph{H}). We see that yearly citations of SBs
decrease rapidly after the peak regardless of their $B$ values. These results
are confirmed also in the WoS dataset, as shown in
Figs.~\ref{fig:beauty-half-life}\emph{I--L}.

\section{Null Models}

To verify that the beauty coefficients cannot be explained by the underlying
citation networks or other well-known mechanisms, we compare the citation
history of each paper as well as the beauty coefficient distribution with
those obtained from some null models. Here we employ two null models on the APS
dataset, namely citation network randomization (NR) and the preferential
attachment mechanism (PA).

The NR procedure starts from the original citation network and carries out a
series of link swapping. The end-point nodes (the papers being cited) of a
randomly selected pair of links (citations) are swapped if: (i) the two links
do not share source or target node; (ii) there are no multiple links after
swapping; and, (iii) the publication year of the cited article is not greater than
that of the citing article after swapping. Performing $Q \cdot E$ switches, where
$E$ is the number of links in the citation network and $Q$ is set to $50$,
yields a transformation of the original citation network into a random directed graph. This
procedure preserves for each paper its number of references (out degree) and
total number of citations (in degree), but destroys the dynamics of yearly
citations.

PA considers as initial network the empirical APS citation network from $1893$
to $1897$ when the first citation occurred; it contains $182$ nodes and 1 link. 
In each following year $t$ until 2009, $n_t$ papers are added at the same time, and
each paper $p$ brings $r_p$ references. $n_t$ is set to the number of APS
papers actually published in year $t$ and each $r_p$ corresponds to the number
of references of one of the papers in such set. As we progressively add papers
to the citation network, the references they contain are addressed to previously
published papers chosen with probability proportional to one plus
the number of citations those papers already have.

\section{Coarse topics of Sleeping Beauties in the APS}

\begin{figure*}
\begin{center}
\includegraphics[trim=0mm 0mm 0mm 0mm, width=0.9\textwidth]{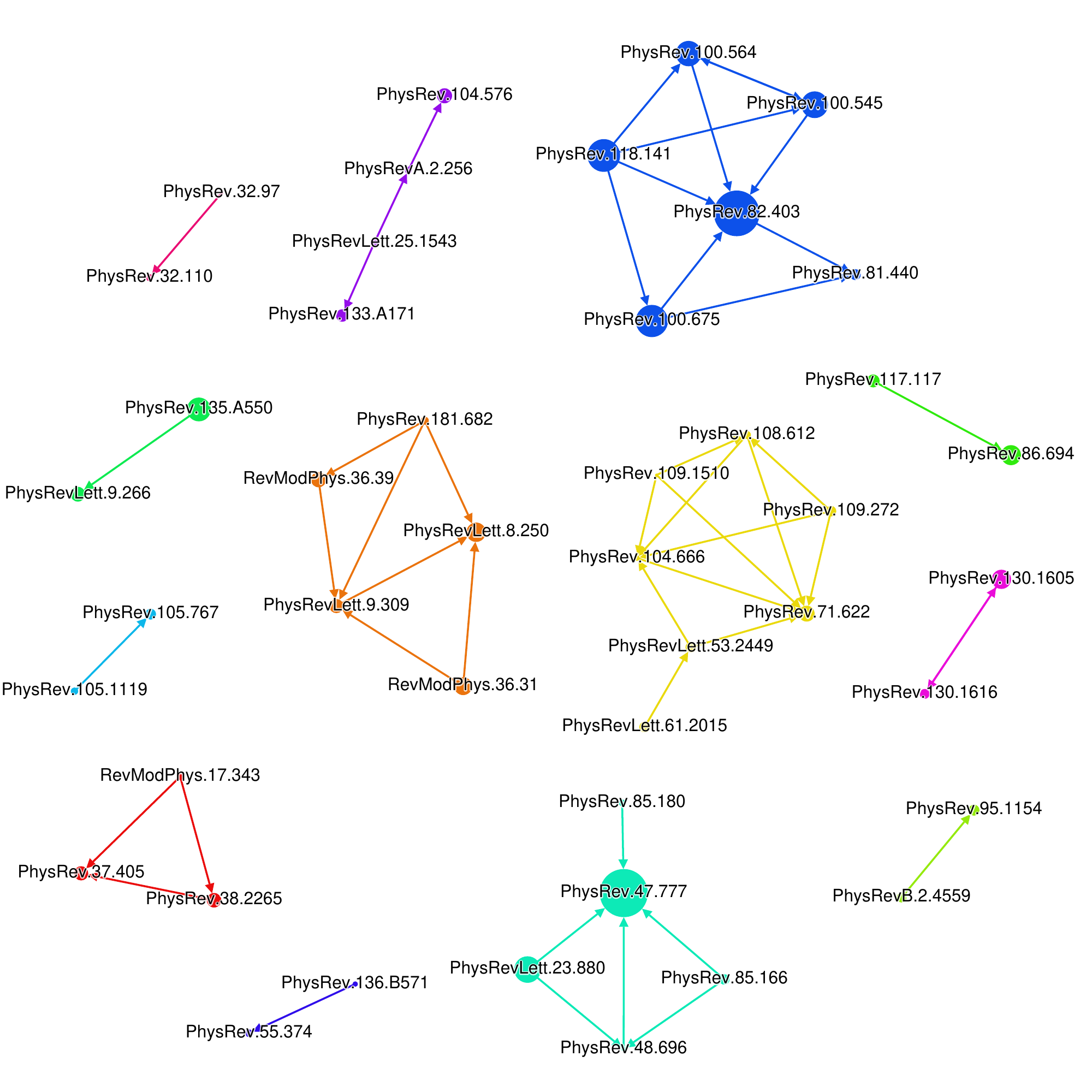}
\caption{\label{fig:beauty-citnet-aps}The citation network of the 100 papers
with highest $B$ values in the APS dataset. Isolated nodes are omitted. The
size of a node is based on its total number of citations.}
\end{center}
\end{figure*}

\begin{figure*}
\begin{center}
\includegraphics[trim=0mm -4mm 0mm 0mm, width=0.45\columnwidth]{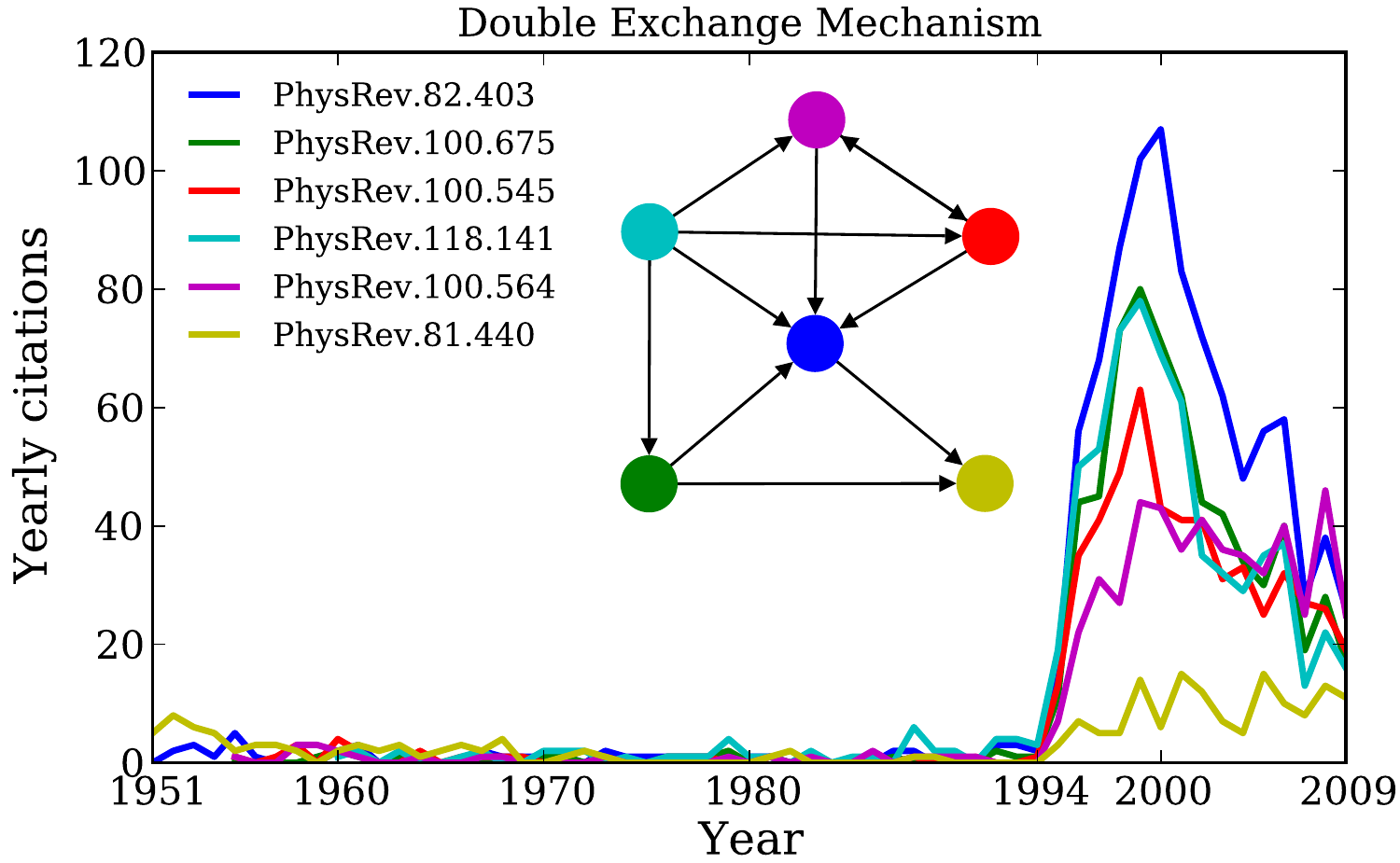}
\includegraphics[trim=0mm -4mm 0mm 0mm, width=0.45\columnwidth]{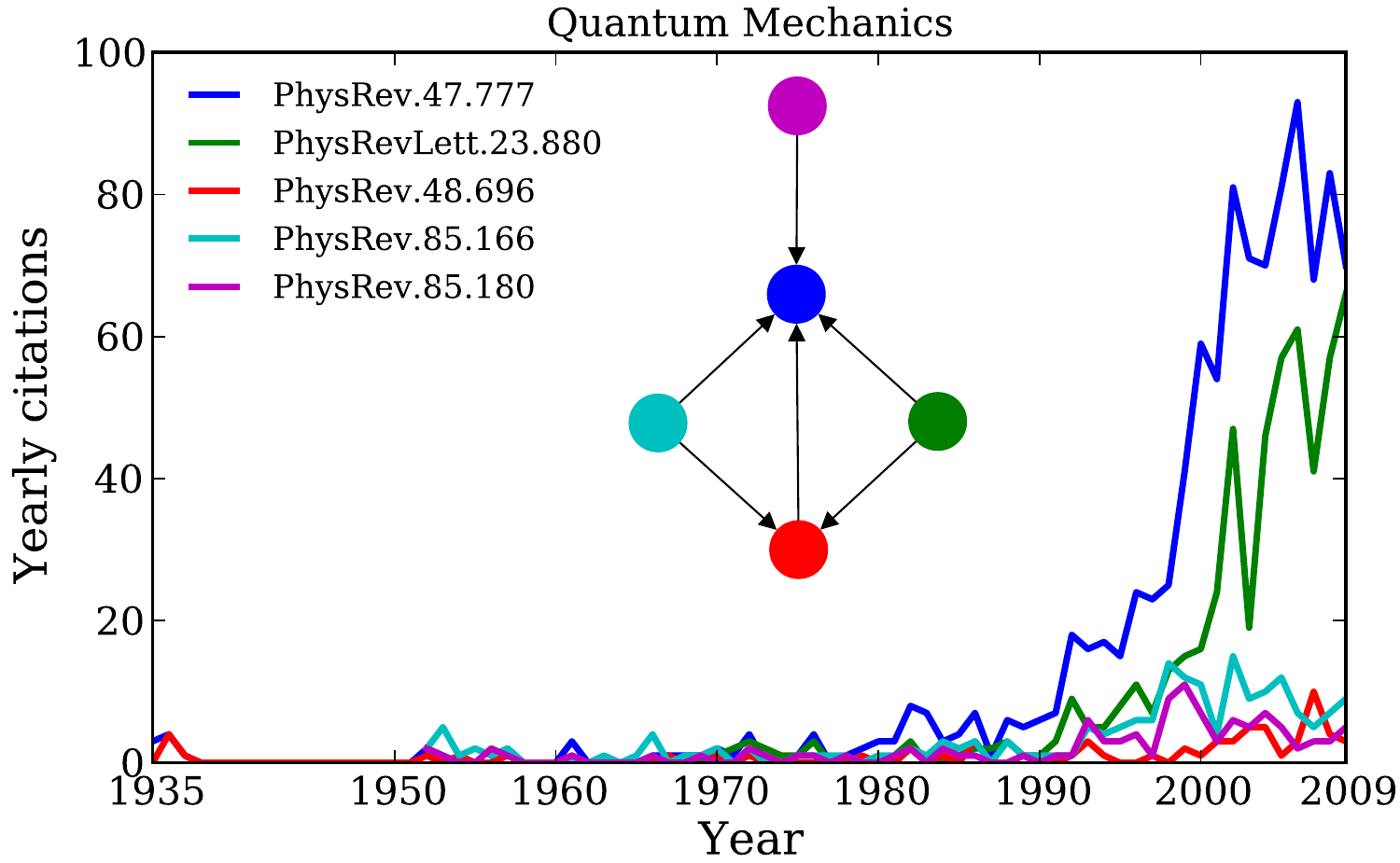}
\includegraphics[trim=0mm 0mm 0mm -4mm, width=0.45\columnwidth]{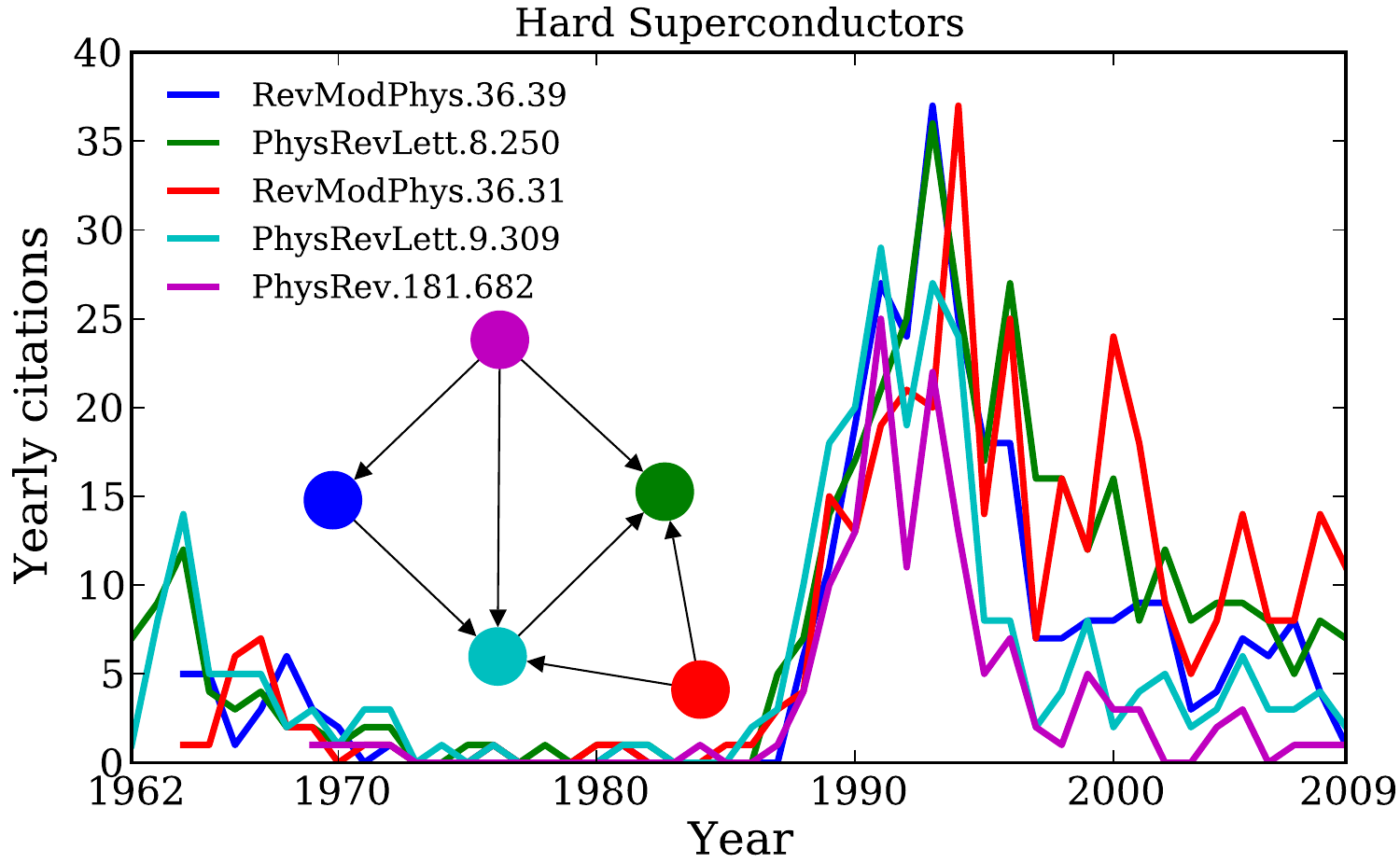}
\includegraphics[trim=0mm 0mm 0mm -4mm, width=0.45\columnwidth]{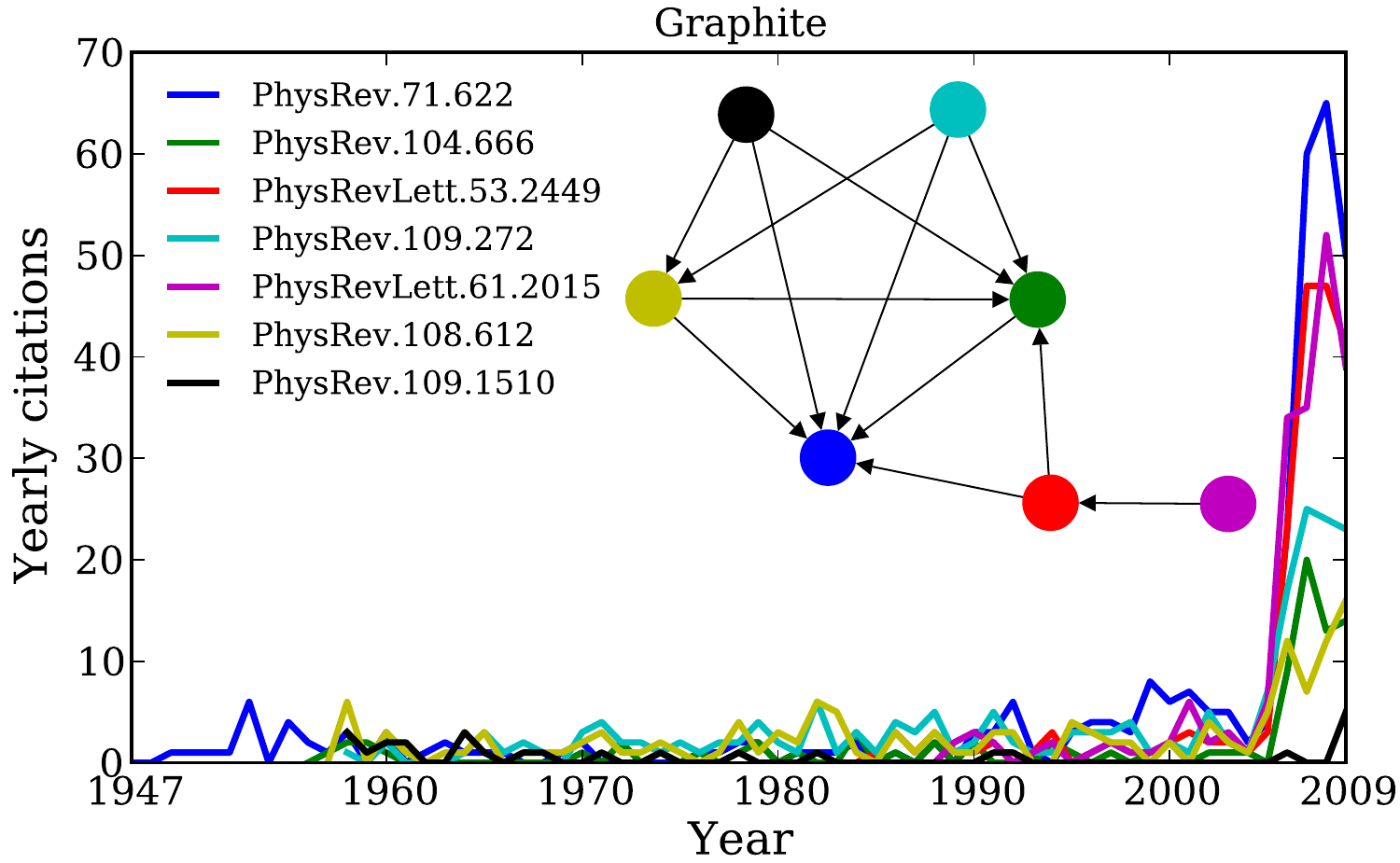}
\caption{\label{fig:beauty-grp}The citation network reveals coarse topics of
Sleeping Beauties. Papers belonging to the same group exhibit similar citation
histories.}
\end{center}
\end{figure*}

Examining the citation relationships between papers with high $B$ values gives
us some coarse topics of Sleeping Beauties. In Fig.~\ref{fig:beauty-citnet-aps}
we present the citation network of the 100 papers with the highest $B$ values
in the APS dataset. Despite many isolated nodes, we observe some (weakly)
connected components. Diving into each component, we find that each one
corresponds to one coarse topic. In Fig.~\ref{fig:beauty-grp}, for instance, we
show the topic of each of the 4 largest components and the citation histories
of its constituent papers. Except for Fig.~\ref{fig:beauty-grp}(b), we observe
that papers belonging to the same group exhibit remarkably similar citation
histories. They are awoken in the same year and exhibit similar up- and
down-going citation patterns. Fig.~\ref{fig:beauty-grp}(a) shows the double
exchange mechanism works. This theory was introduced in 1950s and became
popular in the 1990s. The second group shown in Fig.~\ref{fig:beauty-grp}(b) is
about Quantum Mechanics. The central paper (blue line and blue node), which is
cited by every other paper in the group, is the famous EPR paradox paper by
Einstein, Podolsky, and Rosen. The third group shown in
Fig.~\ref{fig:beauty-grp}(c) is particularly interesting, as it exhibits
complex fluctuations in the citation histories. Finally, the group shown in
Fig.~\ref{fig:beauty-grp}(d) is about graphite and graphene. The central paper
(blue line and blue node) in Fig.~\ref{fig:beauty-grp}(d) is a pioneering work
on the band structure of graphite, foundation of the discovery of graphene, the
subject of the 2010 Nobel Prize in Physics.

\clearpage

\clearpage

\clearpage

\clearpage

\begin{table}
\begin{tabular}{l | c | c | c}
Publication          &   Rank  &  $B$      &  Awakening \\ \hline
PR 40, 749 (1932)    &   45    &  250.79   &  1980 \\
PR 46, 1002 (1934)   &   54    &  237.40   &  1975 \\
PR 47, 777 (1935)    &   2     &  1419.15  &  1987 \\
PR 56, 340 (1939)    &   96    &  174.59   &  1987 \\
PR 82, 403 (1951)    &   1     &  1722.25  &  1994 \\
PR 82, 664 (1951)    &   192   &  122.56   &  2007 \\
PR 100, 545 (1955)   &   4     &  1106.82  &  1994 \\
PR 100, 564 (1955)   &   8     &  670.42   &  1994 \\
PR 100, 675 (1955)   &   3     &  1348.26  &  1994 \\
PR 109, 1492 (1958)  &   147   &  138.63   &  2004 \\
PR 115, 485 (1959)   &   218   &  115.07   &  2001 \\
PR 118, 141 (1960)   &   6     &  841.47   &  1994 \\
\end{tabular}
\caption{\label{tab:redner}Comparison between our results and Redner's results~\cite{Redner-110citation-2005-1}.
The first column lists the 12 \emph{revived classics} in physics detected by Redner's
analysis and arranged in chronological order. From
the second column, we report our results: the rank position according to their
beauty coefficient $B$, the value of $B$, and the awakening year.}
\end{table}

\clearpage

\clearpage

\begingroup
\squeezetable
\begin{table*}
\begin{tabular*}{\hsize}{@{\extracolsep{\fill}}c l l c l}
\hline
$B$ & Author & Title & Pub., awake & Journal\cr \hline
3978  & Pearson, K & On lines and planes of closest fit to systems of points in space & 1901, 2002 & Philos. Mag.\cr \hline
2736  & Wilson, EB & Probable inference, the law of succession, and statistical inference & 1927, 1999 & J. Am. Statist. Assoc.\cr \hline
1909  & Mann, HB & Nonparametric tests against trend & 1945, 2003 & Econometrica\cr \hline
\multirow{2}{*}{1893} & Kaplan, EL; & \multirow{2}{*}{Nonparametric estimation from incomplete observations} & \multirow{2}{*}{1958, 1980} & \multirow{2}{*}{J. Am. Statist. Assoc.}\cr
& Meier, P & & &\cr \hline
\multirow{2}{*}{1760}  & \multirow{2}{*}{Fisher, RA} & On the interpretation of $\chi^2$ from contingency tables, & \multirow{2}{*}{1922, 2006} & \multirow{2}{*}{J. R. Stat. Soc.}\cr
& & and the calculation of $P$ & &\cr \hline
\multirow{2}{*}{1247} & \multirow{2}{*}{Hastings, WK} & Monte-carlo sampling methods using markov chains and & \multirow{2}{*}{1970, 1995} & \multirow{2}{*}{Biometrika}\cr
& & their applications & &\cr \hline
1193  & Metropolis, N & The monte carlo method & 1949, 2004 & J. Am. Statist. Assoc.\cr \hline
1124  & Moran, PAP & Notes on continuous stochastic phenomena & 1950, 1999 & Biometrika\cr \hline
1050  & Lorenz, MO & Methods of measuring the concentration of wealth & 1905, 2005 & J. Am. Statist. Assoc.\cr \hline
985   & Kendall, MG & A new measure of rank correlation & 1938, 2004 & Biometrika\cr
\hline
\end{tabular*}
\caption{\label{tab:beauty-wos-stats}Basic information about the top 10 papers in
Statistics. See Fig.~\ref{fig:beauty-wos-stats} for their citation histories.}
\end{table*}
\endgroup

\clearpage

\begingroup
\squeezetable
\begin{table*}
\begin{tabular*}{\hsize}{@{\extracolsep{\fill}}c l l c l}
\hline
$B$   & Author & Title & Pub., awake & Journal\cr \hline
1215  & Wiener, N & The homogeneous chaos & 1938, 2001 & Amer. J. Math.\cr \hline
1060  & Leray, J & On the movement of a viscous fluid to fill the space & 1934, 1995 & Acta Math.\cr \hline
851   & Pringsheim, A & On the theory of the double infinite numerical orders & 1900, 2005 & Math. Ann.\cr \hline
765   & Jensen, JLWV & On the convex functions and inequalities between mean values & 1906, 2006 & Acta Math.\cr \hline
706   & Mann, WR & Mean value methods in iteration & 1953, 2004 & Proc. Am. Math. Soc.\cr \hline
670   & Halpern, B & Fixed points of nonexpanding maps & 1967, 2004 & Bull. Amer. Math. Soc.\cr \hline
669   & Haar, A & On the theory of orthogonal function systems (first announcement) & 1910, 1988 & Math. Ann.\cr \hline
\multirow{2}{*}{609} & \multirow{2}{*}{Weyl, H} & The asymptotic dispersal law of eigen values of linear partial equations & \multirow{2}{*}{1912, 2002} & \multirow{2}{*}{Math. Ann.}\cr
& & differential (with an application for the theory of cavity radiation) &\cr \hline
\multirow{2}{*}{578} & \multirow{2}{*}{Painleve, P} & About second order and higher order differential equations whose & \multirow{2}{*}{1902, 1990} & \multirow{2}{*}{Acta Math.}\cr
& & general integral is uniform &\cr \hline
\multirow{2}{*}{558} & \multirow{2}{*}{Schmidt, E} & On the theory of linear and non-linear integral equations chapter i & \multirow{2}{*}{1907, 1992} & \multirow{2}{*}{Math. Ann.}\cr
& & development of random functions in specific systems &\cr
\hline
\end{tabular*}
\caption{\label{tab:beauty-wos-math}Basic information about the top 10 papers in
Mathematics. See Fig.~\ref{fig:beauty-wos-math} for their citation histories.}
\end{table*}
\endgroup

\clearpage

\begingroup
\squeezetable
\begin{table*}
\begin{tabular*}{\hsize}{@{\extracolsep{\fill}}c l l c l}
\hline
$B$ & Author & Title & Pub., awake & Journal\cr \hline
1901  & Stroop, JR & Studies of interference in serial verbal reactions & 1935, 1987 & J. Exp. Psychol.\cr \hline
\multirow{2}{*}{1255} & Yerkes, RM; & \multirow{2}{*}{The relation of strength of stimulus to rapidity of habit-formation} & \multirow{2}{*}{1908, 1981} & \multirow{2}{*}{J. Comp. Neurol.}\cr
& Dodson, JD & & &\cr \hline
584   & Zachary, WW & Information flow model for conflict and fission in small groups & 1977, 2005 & J. Anthropol. Res.\cr \hline
563   & Tobler, WR & Computer movie simulating urban growth in Detroit region & 1970, 2003 & Econ. Geogr.\cr \hline
\multirow{2}{*}{545}   & \multirow{2}{*}{Garfield, E} & Citation indexes for science - new dimension in documentation & \multirow{2}{*}{1955, 2000} & \multirow{2}{*}{Science}\cr
& & through association of ideas & &\cr \hline
\multirow{2}{*}{545}   & Heider, F; & \multirow{2}{*}{An experimental study of apparent behavior} & \multirow{2}{*}{1944, 1998} & \multirow{2}{*}{Am. J. Psychol.}\cr
& Simmel, M & & &\cr \hline
521   & Watson, JB & Psychology as the behaviorist views it & 1913, 1968 & Psychol. Rev.\cr \hline
488   & Cohen, J & A coefficient of agreement for nominal scales & 1960, 2009 & Educ. Psychol. Meas.\cr \hline
485   & Maslow, AH & A theory of human motivation & 1943, 1998 & Psychol. Rev.\cr \hline
479   & Glaser, BG & The constant comparative method of qualitative analysis & 1965, 2004 & Social Problems\cr \hline
467   & Todd TW & Age changes in the pubic bone & 1921, 2003 & Am. J. Phys. Anthropol.\cr \hline
460   & Forrester, JW & Industrial dynamics - a major breakthrough for decision makers & 1958, 1993 & HBR\cr \hline
\multirow{2}{*}{453}   & \multirow{2}{*}{Rosenblatt, F} & Perceptron - a probabilistic model for information storage and & \multirow{2}{*}{1958, 2001} & \multirow{2}{*}{Psychol. Rev.}\cr
& & organization in the brain & &\cr \hline
446   & Hotelling, H & Analysis of a complex of statistical variables into principal components & 1933, 1994 & J. Educ. Psychol.\cr \hline
\multirow{2}{*}{428}   & Thorndike, EL; & The influence of improvement in one mental function upon the & \multirow{2}{*}{1901, 1992} & \multirow{2}{*}{Psychol. Rev.}\cr
& Woodworth, RS & of efficiency other functions (I) & &\cr \hline
\multirow{2}{*}{424}   & Holzinger, KJ; & \multirow{2}{*}{The bi-factor method} & \multirow{2}{*}{1937, 2003} & \multirow{2}{*}{Psychometrika}\cr
& Swineford, F & &\cr \hline
\multirow{2}{*}{405}   & Thistlethwaite, DL; & Regression-discontinuity analysis - & \multirow{2}{*}{1960, 2005} & \multirow{2}{*}{J. Educ. Psychol.}\cr
& Campbell, DT & an alternative to the ex-post-facto experiment & &\cr \hline
399   & Horn, JL & A rationale and test for the number of factors in factor-analysis & 1965, 2000 & Psychometrika\cr \hline
375   & Fisher, I & The debt-deflation theory of great depressions & 1933, 2004 & Econometrica\cr \hline
369   & Spitzer, HF & Studies in retention & 1939, 2004 & J. Educ. Psychol.\cr \hline
\multirow{3}{*}{368}   & Linn, BS; & \multirow{3}{*}{Cumulative illness rating scale} & \multirow{3}{*}{1968, 1999} & \multirow{3}{*}{J Am Geriatr Soc.}\cr
& Linn, MW; & & &\cr
& Gurel, L  & & &\cr \hline
358   & Hull, CL & The goal gradient hypothesis and maze learning & 1932, 2001 & Psychol. Rev.\cr \hline
\multirow{2}{*}{356}   & Elftman, H; & \multirow{2}{*}{Chimpanzee and human feet in bipedal walking} & \multirow{2}{*}{1935, 2001} & \multirow{2}{*}{Am. J. Phys. Anthropol.}\cr
& Manter, J & & &\cr \hline
\multirow{2}{*}{349}   & Fornell, C; & Evaluating structural equation models with unobservable variables and & \multirow{2}{*}{1981, 2004} & \multirow{2}{*}{J. Marketing Res.}\cr
& Larcker, DF & measurement error & &\cr \hline
\multirow{2}{*}{343}   & Armstrong, JS; & \multirow{2}{*}{Estimating nonresponse bias in mail surveys} & \multirow{2}{*}{1977, 1998} & \multirow{2}{*}{J. Marketing Res.}\cr
& Overton, TS & & &\cr \hline
342   & Wechsler, H & Toward neutral principles of constitutional-law & 1959, 1986 & Harv. Law Rev.\cr \hline
324   & Cohen, J & Eta-squared and partial eta-squared in fixed factor anova designs & 1973, 2005 & Educ. Psychol. Meas.\cr \hline
324   & Dunlap, K & Reactions to rhythmic stimuli, with attempt to synchronize & 1910, 1995 & Psychol. Rev.\cr \hline
320   & Ellsberg, D & Risk, ambiguity, and the savage axioms & 1961, 2002 & Q. J. Econ.\cr \hline
320   & Lewin, K & Defining the `field at a given time' & 1943, 2006 & Psychol. Rev.\cr
\hline
\end{tabular*}
\caption{\label{tab:beauty-wos-ssh}Basic information about the Sleeping Beauties
in Social Sciences and Humanities among the top $1,000$ in the WoS dataset. See
Fig.~\ref{fig:beauty-wos-ssh-1} and \ref{fig:beauty-wos-ssh-2} for their
citation histories.}
\end{table*}
\endgroup

\clearpage

\clearpage

\clearpage

\clearpage

\clearpage

\clearpage

\begin{figure}
\begin{center}
\includegraphics[trim=0mm 0mm 0mm 0mm, width=0.8\columnwidth]{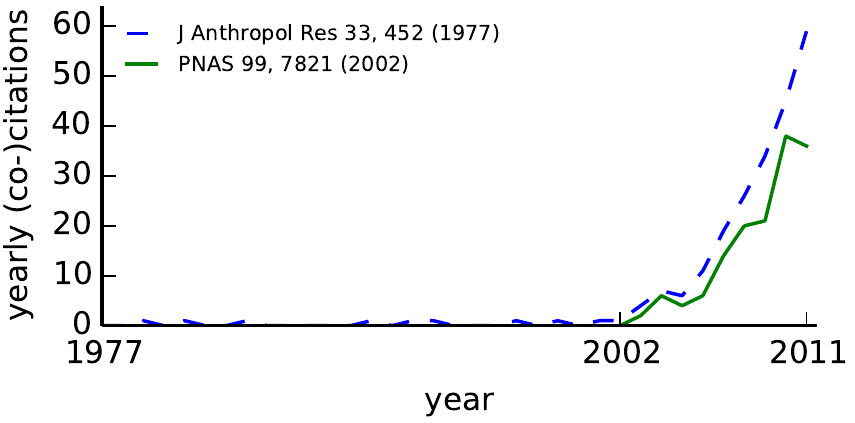}
\caption{\label{fig:zachary}Citation history of the paper
J. Anthropol. Res. \textbf{33}, 452 (1977)~\cite{karate-1977-1}.
The most co-cited paper is
PNAS \textbf{99}, 7821 (2002)~\cite{Girvan-comm-2002-1}.}
\end{center}
\end{figure}

\clearpage

\begin{table}
\begin{tabular}{l | r }
Subject category & Range of $B$\cr \hline
physics, multidisciplinary & [90.56, 5922.97]\cr
chemistry, multidisciplinary & [90.57, 10769.06]\cr
multidisciplinary sciences & [90.54, 3892.49]\cr
mathematics & [90.62, 1215.38]\cr
medicine, general \& internal & [90.58, 1522.30]\cr
physics, applied & [90.63, 3978.42]\cr
surgery & [90.57, 799.65]\cr
chemistry, inorganic \& nuclear & [90.55, 1333.20]\cr
statistics \& probability & [90.56, 2736.18]\cr
mechanics & [90.56, 3978.42]\cr
biology & [90.68, 1247.13]\cr
ecology & [90.60, 1792.29]\cr
physics, condensed matter & [90.58, 3978.42]\cr
biochemistry \& molecular biology & [90.62, 839.22]\cr
astronomy \& astrophysics & [90.56, 984.81]\cr
physics, atomic, molecular \& chemical & [90.60, 774.23]\cr
neurosciences & [90.59, 633.23]\cr
materials science, multidisciplinary & [90.63, 3978.42]\cr
plant sciences & [90.54, 1199.00]\cr
engineering, chemical & [90.60, 2962.53]\cr
\end{tabular}
\caption{\label{tab:b-th}Threshold of $B$ for each of the top $20$ subject
categories producing the top $0.1\%$ SBs in the WoS dataset.}
\end{table}


\clearpage

\begin{thebibliography}{10}

\bibitem{egghe1990introduction}
Egghe L, Rousseau R
\newblock (1990) \emph{Introduction to Informetrics: Quantitative Methods in
  Library, Documentation and Information Science}
\newblock (Elsevier Science, Amsterdam).

\bibitem{Newman-coauthor-04}
Newman MEJ
\newblock (2004) Coauthorship networks and patterns of scientific
  collaboration.
\newblock \emph{Proc Natl Acad Sci USA} 101(suppl 1):5200--5205.

\bibitem{Sun-dynamics-13}
Sun X, Kaur J, Milojevi\'{c} S, Flammini A, Menczer F
\newblock (2013) Social dynamics of science.
\newblock \emph{Sci Rep} 3:1069.

\bibitem{Guimera-team-05}
Guimer\`a R, Uzzi B, Spiro J, Amaral LAN
\newblock (2005) Team assembly mechanisms determine collaboration network
  structure and team performance.
\newblock \emph{Science} 308(5722):697--702.

\bibitem{Wuchty-team-07}
Wuchty S, Jones BF, Uzzi B
\newblock (2007) The increasing dominance of teams in production of knowledge.
\newblock \emph{Science} 316(5827):1036--1039.

\bibitem{Jones-team-08}
Jones BF, Wuchty S, Uzzi B
\newblock (2008) Multi-university research teams: Shifting impact, geography,
  and stratification in science.
\newblock \emph{Science} 322(5905):1259--1262.

\bibitem{Milojevic-team-14}
Milojevi\'c S
\newblock (2014) Principles of scientific research team formation and
  evolution.
\newblock \emph{Proc Natl Acad Sci USA} 111(11):3984--3989.

\bibitem{Radicchi-impact-08}
Radicchi F, Fortunato S, Castellano C
\newblock (2008) Universality of citation distributions: Toward an objective
  measure of scientific impact.
\newblock \emph{Proc Natl Acad Sci USA} 105(45):17268--17272.

\bibitem{Wang-impact-13}
Wang D, Song C, Barab\'asi AL
\newblock (2013) Quantifying long-term scientific impact.
\newblock \emph{Science} 342(6154):127--132.

\bibitem{Uzzi-comb-13}
Uzzi B, Mukherjee S, Stringer M, Jones B
\newblock (2013) Atypical combinations and scientific impact.
\newblock \emph{Science} 342(6157):468--472.

\bibitem{hirsch2005index}
Hirsch JE
\newblock (2005) An index to quantify an individual's scientific research
  output.
\newblock \emph{Proc Natl Acad Sci USA} 102(46):16569--16572.

\bibitem{kinney2007national}
Kinney A
\newblock (2007) National scientific facilities and their science impact on
  nonbiomedical research.
\newblock \emph{Proc Natl Acad Sci USA} 104(46):17943--17947.

\bibitem{davis1984faculty}
Davis P, Papanek GF
\newblock (1984) Faculty ratings of major economics departments by citations.
\newblock \emph{Am Econ Rev} 74(1):225--230.

\bibitem{bornmann2006selecting}
Bornmann L, Daniel HD
\newblock (2006) Selecting scientific excellence through committee peer
  review-a citation analysis of publications previously published to approval
  or rejection of post-doctoral research fellowship applicants.
\newblock \emph{Scientometrics} 68(3):427--440.

\bibitem{liu2005academic}
Liu NC, Cheng Y
\newblock (2005) The academic ranking of world universities.
\newblock \emph{High Educ Eur} 30:127--136.

\bibitem{sarigol2014predicting}
Sarig{\"o}l E, Pfitzner R, Scholtes I, Garas A, Schweitzer F
\newblock (2014) Predicting scientific success based on coauthorship networks.
\newblock \emph{EPJ Data Science} 3(1):9.

\bibitem{Petersen-reputation-14}
Petersen AM, {et~al.}
\newblock (2014) Reputation and impact in academic careers.
\newblock \emph{Proc Natl Acad Sci USA} 111(43):15316--15321.

\bibitem{Price-cumuadv-76}
de~Solla~Price DJ
\newblock (1976) A general theory of bibliometric and other cumulative
  advantage processes.
\newblock \emph{J Am Soc Inf Sci}
  27(5):292--306.

\bibitem{Barabasi-pa-99}
Barab\'asi AL, Albert R
\newblock (1999) Emergence of scaling in random networks.
\newblock \emph{Science} 286(5439):509--512.

\bibitem{albert2002statistical}
Albert R, Barab{\'a}si AL
\newblock (2002) Statistical mechanics of complex networks.
\newblock \emph{Rev Mod Phys} 74(1):47--97.

\bibitem{boccaletti2006complex}
Boccaletti S, Latora V, Moreno Y, Chavez M, Hwang DU
\newblock (2006) Complex networks: Structure and dynamics.
\newblock \emph{Phys Rep} 424(4--5):175--308.

\bibitem{Krapivsky-redirect-01}
Krapivsky PL, Redner S
\newblock (2001) Organization of growing random networks.
\newblock \emph{Phys Rev E Stat Nonlin Soft Matter Phys} 63(6 Pt 2):066123.

\bibitem{newman2009first}
Newman MEJ
\newblock (2009) The first-mover advantage in scientific publication.
\newblock \emph{EPL} 86(6):68001.

\bibitem{hajra2004phase}
Hajra KB, Sen P
\newblock (2004) Phase transitions in an aging network.
\newblock \emph{Phys Rev E Stat Nonlin Soft Matter Phys} 70(5 Pt 2):056103.

\bibitem{hajra2005aging}
Hajra KB, Sen P
\newblock (2005) Aging in citation networks.
\newblock \emph{Physica A} 346(1--2):44--48.

\bibitem{hajra2006modelling}
Hajra KB, Sen P
\newblock (2006) Modelling aging characteristics in citation networks.
\newblock \emph{Physica A} 368(2):575--582.

\bibitem{wang2008measuring}
Wang M, Yu G, Yu D
\newblock (2008) Measuring the preferential attachment mechanism in citation
  networks.
\newblock \emph{Physica A} 387(18):4692--4698.

\bibitem{dorogovtsev2000evolution}
Dorogovtsev SN, Mendes JFF
\newblock (2000) Evolution of networks with aging of sites.
\newblock \emph{Phys Rev E Stat Nonlin Soft Matter Phys} 62:1842.

\bibitem{dorogovtsev2001scaling}
Dorogovtsev SN, Mendes JF
\newblock (2001) Scaling properties of scale-free evolving networks: Continuous
  approach.
\newblock \emph{Phys Rev E Stat Nonlin Soft Matter Phys} 63(5 Pt 2):056125.

\bibitem{zhu2003effect}
Zhu H, Wang X, Zhu JY
\newblock (2003) Effect of aging on network structure.
\newblock \emph{Phys Rev E Stat Nonlin Soft Matter Phys} 68(5 Pt 2):056121.

\bibitem{Garfield-beauty-80}
Garfield E
\newblock (1980) Premature discovery or delayed recognition---why?
\newblock \emph{Current Contents} 21:5--10.

\bibitem{Garfield-beauty-89}
Garfield E
\newblock (1989) Delayed recognition in scientific discovery: Citation
  frequency analysis aids the search for case histories.
\newblock \emph{Current Contents} 23:3--9.

\bibitem{Garfield-beautyb-89}
Garfield E
\newblock (1989) More delayed recognition. Part 1. Examples from the genetics
  of color blindness, the entropy of short-term memory, phosphoinositides, and
  polymer rheology.
\newblock \emph{Current Contents} 38:3--8.

\bibitem{Garfield-beauty-90}
Garfield E
\newblock (1990) More delayed recognition. Part 2. From inhibin to scanning
  electron microscopy.
\newblock \emph{Current Contents} 9:3--9.

\bibitem{Glanzel-beauty-03}
Gl\"anzel W, Schlemmer B, Thijs B
\newblock (2003) Better late than never? On the chance to become highly cited
  only beyond the standard bibliometric time horizon.
\newblock \emph{Scientometrics} 58(3):571--586.

\bibitem{Raan-Beauty-04}
van Raan AFJ
\newblock (2004) Sleeping Beauties in science.
\newblock \emph{Scientometrics} 59(3):467--472.

\bibitem{Redner-110citation-2005}
Redner S
\newblock (2005) Citation statistics from 110 years of physical review.
\newblock \emph{Phys Today} 58(6):49--54.

\bibitem{Bornmann-percentile-2013}
Bornmann L, Leydesdorff L, Wang J
\newblock (2013) Which percentile-based approach should be preferred for
  calculating normalized citation impact values? An empirical comparison of
  five approaches including a newly developed citation-rank approach (p100).
\newblock \emph{J Informetrics} 7(4):933--944.

\bibitem{Bornmann-improve-2014}
Bornmann L, Leydesdorff L, Wang J
\newblock (2014) How to improve the prediction based on citation impact
  percentiles for years shortly after the publication date?
\newblock \emph{J Informetrics} 8(1):175--180.

\bibitem{Wang-window-2013}
Wang J
\newblock (2013) Citation time window choice for research impact evaluation.
\newblock \emph{Scientometrics} 94(3):851--872.

\bibitem{Glanzel-beauty-04}
Gl\"anzel W, Garfield E
\newblock (2004) The myth of delayed recognition.
\newblock \emph{The Scientist} 18:8--9.

\bibitem{marx2010reference}
Marx W, Bornmann L, Cardona M
\newblock (2010) Reference standards and reference multipliers for the
  comparison of the citation impact of papers published in different time
  periods.
\newblock \emph{J Am Soc Info Sci Technol} 61(10):2061--2069.

\bibitem{Garfield-sci-1955}
Garfield E
\newblock (1955) Citation indexes for science: A new dimension in documentation
  through association of ideas.
\newblock \emph{Science} 122(3159):108--111.

\bibitem{marx2014shockley}
Marx W
\newblock (2014) The Shockley-Queisser paper--a notable example of a scientific
  sleeping beauty.
\newblock \emph{Annalen der Physik} 526(5-6):A41--A45.

\bibitem{kleinberg1999authoritative}
Kleinberg JM
\newblock (1999) Authoritative sources in a hyperlinked environment.
\newblock \emph{J ACM} 46(5):604--632.

\bibitem{seglen1997impact}
Seglen PO
\newblock (1997) Why the impact factor of journals should not be used for
  evaluating research.
\newblock \emph{BMJ} 314(7079):497.

\bibitem{Zachary-karate-1977}
Zachary WW
\newblock (1977) An information flow model for conflict and fission in small
  groups.
\newblock \emph{J Anthropol Res} 33(4):452--473.

\bibitem{Girvan-comm-2002}
Girvan M, Newman MEJ
\newblock (2002) Community structure in social and biological networks.
\newblock \emph{Proc Natl Acad Sci USA}
  99(12):7821--7826.

\bibitem{PhysRev.95.1154}
Karplus R, Luttinger J
\newblock (1954) Hall effect in ferromagnetics.
\newblock \emph{Phys Rev} 95(5):1154--1160.

\bibitem{PhysRev.82.403}
Zener C
\newblock (1951) Interaction between the $d$-shells in the transition metals.
  ii. ferromagnetic compounds of manganese with perovskite structure.
\newblock \emph{Phys Rev} 82(3):403--405.

\bibitem{PhysRevB.58.12547}
Molina M
\newblock (1998) Transport of localized and extended excitations in a nonlinear
  anderson model.
\newblock \emph{Phys Rev B} 58(19):12547--12550.

\bibitem{PhysRev.78.294}
Nordheim L
\newblock (1950) $\beta$-decay and the nuclear shell model.
\newblock \emph{Phys Rev} 78(3):294.

\bibitem{PhysRevLett.62.324}
Metzner W, Vollhardt D
\newblock (1989) Correlated lattice fermions in $d=\infty$ dimensions.
\newblock \emph{Phys Rev Lett} 62(3):324--327.

\bibitem{Clauset-powerlaw-09}
Clauset A, Shalizi CR, Newman MEJ
\newblock (2009) Power-law distributions in empirical data.
\newblock \emph{SIAM Rev} 51(4):661--703.

\bibitem{Garfield-jama-2006}
Garfield E
\newblock (2006) The history and meaning of the journal impact factor.
\newblock \emph{JAMA} 295(1):90--93.

\bibitem{Garfield-science-1972}
Garfield E
\newblock (1972) Citation analysis as a tool in journal evaluation.
\newblock \emph{Science} 178(4060):471--479.

\bibitem{Garfield-cmaj-1999}
Garfield E
\newblock (1999) Journal impact factor: A brief review.
\newblock \emph{Can Med Assoc J} 161(8):979--980.

\end{thebibliography}

\clearpage

\begin{thebibliography}{10}

\bibitem{PhysRev.100.675}
P.~Anderson and H.~Hasegawa.
\newblock Considerations on double exchange.
\newblock {\em Phys. Rev.}, 100:675--681, Oct 1955.

\bibitem{PhysRev.118.141}
P.~de~Gennes.
\newblock Effects of double exchange in magnetic crystals.
\newblock {\em Phys. Rev.}, 118:141--154, Apr 1960.

\bibitem{PhysRev.100.580}
G.~Dresselhaus.
\newblock Spin-orbit coupling effects in zinc blende structures.
\newblock {\em Phys. Rev.}, 100:580--586, Oct 1955.

\bibitem{PhysRev.47.777}
A.~Einstein, B.~Podolsky, and N.~Rosen.
\newblock Can quantum-mechanical description of physical reality be considered
  complete?
\newblock {\em Phys. Rev.}, 47:777--780, May 1935.

\bibitem{PhysRev.135.A550}
P.~Fulde and R.~Ferrell.
\newblock Superconductivity in a strong spin-exchange field.
\newblock {\em Phys. Rev.}, 135:A550--A563, Aug 1964.

\bibitem{Girvan-comm-2002-1}
M.~Girvan and M.~E.~J. Newman.
\newblock Community structure in social and biological networks.
\newblock {\em Proceedings of the National Academy of Sciences},
  99(12):7821--7826, 2002.

\bibitem{PhysRev.100.564}
J.~Goodenough.
\newblock Theory of the role of covalence in the perovskite-type manganites
  $[\mathrm{La}, m(\mathrm{II})]\mathrm{Mn}{\mathrm{o}}_{3}$.
\newblock {\em Phys. Rev.}, 100:564--573, Oct 1955.

\bibitem{Redner-110citation-2005-1}
S.~Redner.
\newblock Citation statistics from 110 years of physical review.
\newblock {\em Physics Today}, 58(6):49--54, 2005.

\bibitem{PhysRev.71.622}
P.~Wallace.
\newblock The band theory of graphite.
\newblock {\em Phys. Rev.}, 71:622--634, May 1947.

\bibitem{PhysRev.100.545}
E.~Wollan and W.~Koehler.
\newblock Neutron diffraction study of the magnetic properties of the series of
  perovskite-type compounds $[(1-x)\mathrm{La},
  x\mathrm{Ca}]\mathrm{Mn}{\mathrm{o}}_{3}$.
\newblock {\em Phys. Rev.}, 100:545--563, Oct 1955.

\bibitem{karate-1977-1}
W.~W. Zachary.
\newblock An information flow model for conflict and fission in small groups.
\newblock {\em Journal of Anthropological Research}, 33(4):452--473, 1977.

\bibitem{PhysRev.82.403-1}
C.~Zener.
\newblock Interaction between the $d$-shells in the transition metals. ii.
  ferromagnetic compounds of manganese with perovskite structure.
\newblock {\em Phys. Rev.}, 82:403--405, May 1951.

\end{thebibliography}
\end{document}